\documentclass[lettersize,journal]{IEEEtran}
\usepackage{amsmath,amsfonts,amssymb}
\usepackage{graphicx}
\usepackage{algorithmicx}
\usepackage[ruled,linesnumbered]{algorithm2e}
\usepackage{algpseudocode}
\usepackage{textcomp}
\usepackage{makecell}
\usepackage{multirow,multicol}
\usepackage[colorlinks]{hyperref}
\hypersetup{citecolor=blue}
\usepackage[table]{xcolor}
\usepackage{threeparttable}
\usepackage{booktabs}
\usepackage{float}
\usepackage{balance}
\begin{document}

\title{Joint Activity Detection and Channel Estimation For Fluid Antenna System Exploiting Geographical and Angular Information}
\author{Zhentian Zhang,~\IEEEmembership{Graduate Student Member,~IEEE}, Jian Dang,~\IEEEmembership{Senior Member,~IEEE}, David Morales-Jimenez,~\IEEEmembership{Senior Member,~IEEE}, Hao Jiang,~\IEEEmembership{Senior Member,~IEEE}, Zaichen Zhang,~\IEEEmembership{Senior Member,~IEEE}, Christos Masouros,~\IEEEmembership{Fellow,~IEEE,} and 
	Chan-Byoung Chae,~\IEEEmembership{Fellow,~IEEE}
	\vspace{-5mm}
	
	\thanks{ }
	\thanks{The work of Zhentian Zhang, Jian Dang and Zaichen Zhang is supported in part by the Fundamental Research Funds for the Central Universities (2242022k60001) , Jiangsu NSF project (No. BK20252003),  the Key Laboratory of Intelligent Support Technology for Complex Environments, Ministry of Education, Nanjing University of Information Science and Technology (No. B2202402). The work of David Morales-Jimenez is supported in part by the State Research Agency (AEI) of Spain and the European Social Fund under grant RYC2020-030536-I and by MICIU/AEI/10.13039/501100011033 under grant PID2023-149975OB-I00 (COSTUME). The work of Hao Jiang is supported in part by the National Natural Science Foundation of China (NSFC) projects (No. 62471238). The work of Chan-Byoung Chae was supported in part by the Korea government under IITP/NRF Grants 2022-0-00704 and 2022R1A5A1027646.}
	\thanks{Zhentian Zhang, Jian Dang, Zaichen Zhang are with the National Mobile Communications Research Laboratory, Frontiers Science Center for Mobile Information Communication and Security, Southeast University, Nanjing, 210096, China. Jian Dang is also with Key Laboratory of Intelligent Support Technology for Complex Environments, Ministry of Education, Nanjing University of Information Science and Technology, Nanjing 210044, China. Jian Dang, Zaichen Zhang are also with the Purple Mountain Laboratories, Nanjing 211111, China (e-mail: \{zhangzhentian, dangjian, zczhang\}@seu.edu.cn).}
	\thanks{David Morales-Jimenez is with the Department of Signal Theory, Networking and Communications, University of Granada, Granada 18071, Spain (e-mail: dmorales@ugr.es).}
	\thanks{Hao Jiang is with the School of Artificial Intelligence, Nanjing University of Information Science and Technology, Nanjing 210044, China, and also with the National Mobile Communications Research Laboratory, Southeast University, Nanjing 210096, China (e-mail: jianghao@nuist.edu.cn).}
	\thanks{Christos Masouros are with the Department of Electronic and Electrical Engineering, University College London, Torrington Place, WC1E 7JE, United Kingdom (e-mail: c.masouros@ucl.ac.uk).}
	\thanks{Chan-Byoung Chae is with the School of Integrated Technology, Seoul 03722 Korea (e-mail: cbchae@yonsei.ac.kr).}
	\thanks{Corresponding authors: J. Dang (dangjian@seu.edu.cn)}
	}



\maketitle

\begin{abstract}
The fluid antenna system (FAS) refers to a family of reconfigurable antenna technologies that provide substantial spatial gains within a compact, predefined small space, thereby offering extensive degrees of freedom in the physical layer for future communication networks. The acquisition of channel state information (CSI) is critical, as it determines the placement of ports/antennas, which directly impacts FAS-based optimization. Although various channel estimation methods have been developed, significant flaws persist. For instance, the performance of greedy-based algorithms is heavily influenced by signal assumptions, and current model-free methods are infeasible due to prohibitively high computational complexity issue. Consequently, there is a pressing need for a well-balanced solution that exhibits flexibility, feasibility, and low complexity to support massive connectivity in FAS. In this work, we propose methods based on approximate message passing (AMP) integrated with adaptive expectation maximization (EM). The EM-AMP framework uniquely enables efficient large matrix computations with adaptive learning capabilities, independent of prior knowledge of the model or parameters within potential distributions, making it a robust candidate for FAS networks. We introduce two variants of the EM-AMP framework that leverage geographical and angular features in a FAS network. These proposed algorithms demonstrate improved estimation precision, fast convergence, and low computational complexity in large activity regions. Additionally, we analytically elucidate the reasons behind the inherent performance floor of greedy-based methods and highlight the critical role of angular information in algorithm design. Extensive numerical results validate the promising efficacy of the proposed algorithm designs and the derived analytical findings.
\end{abstract}


\section{Introduction}
\subsection{Background and Related Work}
Massive communication builds upon the concept of massive machine-type communication (mMTC) from International Mobile Telecommunications (IMT)-2020, enabling the connectivity of a vast number of long-lasting devices or sensors for diverse Internet-of-Things (IoT) applications \cite{mMTC1,mMTC2}. Among all potential enablers for unprecedented massive connectivity, the fluid antenna system (FAS) \cite{FAS1,fbl_fas,fas_port}, as a novel series of reconfigurable antenna technologies \cite{fas_tur2,fas_tur3, FAS1.1,FAS1.2}, stands out by providing extensive degrees of freedom (DOF) at the physical layer. An FAS device can fully utilize the spatial domain of a predefined antenna, thereby achieving significant spatial gains within a compact space \cite{FAS1.4,FAS1.3}.

The FAS particularly distinguishes itself due to its unique channel response envelope behavior and the inherent diversity stemming from the correlation among ports at different locations \cite{FAS_channel 1,FAS_channel 2,FAS_channel 3,FAS_channel 4,FAS_channel 5}. Notably, the radical channel response fluctuations within a confined spatial domain create favorable interference gaps among users, yielding significant spatial gains for massive connectivity and thereby transforming the multiple access framework, namely, fluid antenna multiple access (FAMA) \cite{FAMA1,FAMA2,FAMA3,FAMA4}. Moreover, FAS seamlessly integrates with other advanced technologies, such as reconfigurable intelligent surface (RIS) \cite{FAS-RIS1,FAS-RIS2,FAS-RIS3,FAS-RIS4,FAS-RIS5} and integrated sensing and communications (ISAC) \cite{FAS-ISAC1,FAS-ISAC2,FAS-ISAC3}, manifesting incredible versatility.

Although the potential DOF within FAS could enable robust communication in a channel state information-free (CSI-free) or CSI-less manner \cite{CSI-free1,CSI-free2}, CSI acquisition remains critical for essential tasks such as antenna placement \cite{CE1}. Within a coherence time, acquiring CSI for ports at different locations requires a substantial number of pilots. Moreover, due to the hardware constraints of FAS receiver, only a limited number of ports/antennas can be connected to radio frequency (RF) chains for pilot reception within the coherence time, which significantly complicates channel estimation. To address this challenge, various approaches have been explored. In \cite{FAS_channel 0.1}, a least square (LS) solution is investigated under uniform port selection, combined with covariance-gradient optimization for active port gap selection. However, this approach assumes perfect prior knowledge, including angle of arrivals (AoAs) and noise variance. To reduce dependency on priors, \cite{FAS_channel 0.2} proposes an AoA-codebook-based solution and examines the impact of position gaps on the minimum mean square error (MMSE) estimation. Additionally, \cite{CE2} quantifies the minimum number of estimated channels and the total number of pilot symbols required for efficient channel reconstruction within a given space. 

Self-adaptive algorithm designs have become appealing features due to their capability of learning vital parameters. \cite{CE4} introduces a sparse Bayesian learning (SBL) framework into FAS, although it relies on an accurate signal prior distribution and suffers from quadratic computational complexity. To circumvent the requirement for a known signal model, \cite{CE3} models FAS channels as a stochastic process, where uncertainty is successively reduced through kernel-based sampling and regression, thereby providing a highly versatile, distribution-unaware detection solution. However, this approach incurs prohibitively high computational complexity in the cubic order with respect to pilot length and the number of active ports, making it less suitable for scenarios involving large array systems or substantial pilot overhead under massive connectivity.

Greedy-based algorithms may underperform in suboptimal conditions but they remain practically and feasibly deployable. \cite{CE6} explores irregular antenna shaping with a linear MMSE (LMMSE) estimator, and \cite{CE5} introduces a low-sample-size sparse channel reconstruction (L3SCR) method, achieving precise CSI with minimal hardware switching and pilot overhead. Nevertheless, its performance is inferior to the LS solution when a relatively large number of active receiving ports are used \cite[Fig.~2]{CE5} or in high signal-to-noise ratio (SNR) regions \cite[Fig.~4]{CE5}.

\subsection{Motivation and Contributions}\label{sec.1-B}
In this paper, we introduce the approximate message passing (AMP) framework \cite{AMP1} for CSI acquisition in FAS, an approach that, {\em to the best of our knowledge, has not been previously investigated for FAS}. Notably, AMP inherently excels in signal processing under unknown models \cite{AMP2}. Furthermore, its computational flow can be effectively designed in a semi-blind manner based on the potential signal model, with all critical parameters accurately learned from noisy observations \cite{EM-AMP1}. Additionally, as an iterative algorithm framework, AMP is particularly well-suited for large-scale matrix computations, offering computational complexity of linear order. This efficiency makes the AMP framework a strong candidate for massive access applications such as CSI acquisition \cite{EM-AMP2,EM-AMP3}, multi-user detection \cite{AMP3,AMP4,AMP5}, and related tasks. Our contributions are summarized as follows.

\begin{itemize}
	\item Initially, we provide insights into an seemingly inherent performance floor phenomenon observed in greedy correlation-based algorithms, as illustrated in \cite[Fig.~12, Fig.~13]{FAS1} and \cite[Fig.~2, Fig.~4]{CE5}. Analytically, we elucidate why the estimation performance converges to a fixed level when a relatively large number of active receiving ports are employed \cite[Fig.~2]{CE5} or in high SNR regions \cite[Fig.~4]{CE5}. Our findings reveal that {\em the estimation performance is directly influenced by the ability to accurately estimate the variance of the potential signal model}, thereby providing valuable guidance for algorithm design.
	
	\item Building on prior findings, we first provide a detailed explanation of how to implement the AMP framework with expectation maximization (EM) for CSI acquisition in FAS under unknown parameters. Subsequently, we derive novel update rules by leveraging the {\em geographical features} of the FAS network. These update rules enhance the estimation on the prior probability density function (PDF) variance, thereby improving estimation precision. Crucially, the newly derived update rules introduce minimal additional computational complexity while exhibiting significantly faster convergence speed compared to the original framework.
	
	\item Furthermore, we propose an EM-AMP approach that leverages {\em angular information} inherent in noisy observations, significantly enhancing estimation performance and convergence speed. Unlike LS-based or greedy-based methods, the improved angle resolution from an increased number of active ports further strengthens the estimation capability. Additionally, we provide empirical methods to utilize angular information within the EM-AMP framework and demonstrate the necessity and significance of angular information utility by deriving the approximation of estimation error in/not in aware of angular information and verify the analytical results by providing empirical results.
	
	\item Moreover, unlike the SBL method in \cite{CE4} and the kernel-based learning method in \cite{CE3}, whose computational complexities are respectively in the quadratic and cubic orders with respect to the matrix size, the proposed framework requires only linear-order computational complexity.
	
	\item To facilitate understanding of the updates of the posteriors and priors in \eqref{eq:17} and \eqref{eq:20}, respectively, we provide detailed step-by-step derivations for both the AMP and EM procedures. The complete derivations are available at {\em https://github.com/BrooklynSEUPHD/Supplementary-Material-Step-wise-Derivations-on-EM-AMP.git}.
\end{itemize}

The remainder of this paper is organized as follows. In Section~\ref{system model}, the signal models are described in detail. In Section~\ref{inherent_problem}, we analytically explain the inherent performance floor phenomenon of greedy-based algorithms. In Section~\ref{proposed}, we elaborate on the proposed EM-AMP algorithms, which leverage geographical and angular information, providing both practical update rules and analytical proofs. In Section~\ref{Numerical Results}, we present extensive numerical results, and finally, conclusions are drawn in Section~\ref{conclusion}.

 \textit{Notations:} vectors and matrices are represented by bold lowercase and uppercase letters, respectively. The sets of real and complex numbers are denoted by $\mathbb{R}$ and $\mathbb{C}$.  Sets are written in calligraphy style, e.g., $\mathcal{A}$. The element in the $m$-th row and $n$-th column of a matrix is denoted by $\mathbf{A}\left[m,n\right]$. The operations $\left(\cdot\right)^{\mathrm{T}}$, $\left(\cdot\right)^{\mathrm{H}}$ denote transpose and Hermitian transpose, respectively. With a little abuse of notation, the PDF of a complex Gaussian random variable $x$ with mean $\mu$ and variance $\phi$ is denoted as $\mathcal{CN}(x;\mu,\phi)=\frac{1}{\pi\phi}e^{-\frac{|x-\mu|^2}{\phi}}$. For a real Gaussian random variable $x$, the corresponding PDF is $\mathcal{N}(x;\mu,\phi)=\frac{1}{\sqrt{2\pi\phi}}e^{-\frac{(x-\mu)^{2}}{2\phi}}$. The operations $|\cdot|$ and $\|\cdot\|_2^2$ denote modulus and $l_2$-norm, respectively.
\begin{figure}[t!]
	\centering
	\includegraphics[width=\columnwidth]{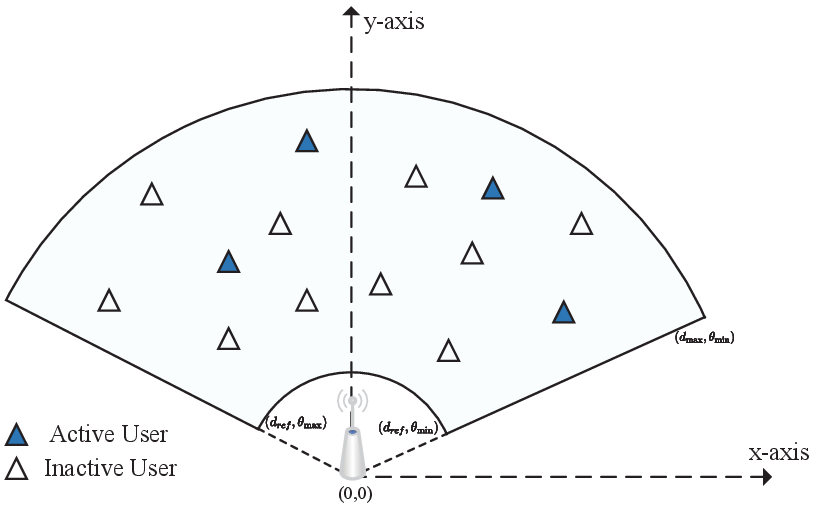}
	\caption{Illustration of system model, where the receiver is deemed as a linear array with a service area in a circle sector.}
	\label{fig:system_model}
\end{figure}
\section{System Model}\label{system model}
\subsection{Signal Model}
Consider a grant-free uplink transmission scenario in which a base station (BS) equipped with a $W$-length fluid antenna serves $K$ potential single-antenna users. The transmission resources are organized into frames, with each frame containing $G$ single-carrier modulated symbols that serve as pilot signals for uplink channel training. It is assumed that the traffic is sporadic, meaning that only $K_a$ out of the $K$ users are active in each frame, while the remaining users remain idle. Each user is allocated with an unique pilot codeword and let $\mathbf{a}_k\in \mathbb{C}^{G\times 1}$ denote the $k$-th user pilot and the BS restores all codewords as a pilot codebook $\mathbf{A}=\left[\mathbf{a}_1,\ldots,\mathbf{a}_K\right]\in \mathbb{C}^{G\times K}$. For ease of description, the energy level of any pilot codeword is normalized as unit power, i.e., $\|\mathbf{a}_k\|_2^2=1$. Due to the limited number of radio frequence (RF) chains at the BS side, only $N_o$ ports will be activated within a length-fixed fluid antenna at the BS. Let $\mathbf{h}_k\in \mathbb{C}^{1\times N_o}$ denote the channel coefficients between the $k$-th user and the BS.

Omitting asynchronous transmission, the overlapped signals at the BS side can be written as:
\begin{equation}\label{eq:1}
	\mathbf{Y} = \sum_{k=1}^{K}\alpha_k\mathbf{a}_k\mathbf{h}_k+\mathbf{Z},
\end{equation}
where $\mathbf{Y}\in\mathbb{C}^{G\times N_o}$ is the received signal, constant $\alpha_k$ indicates the activity of $k$-th pilot, i.e., if $\alpha_k=1$, $\mathbf{a}_k$ is active, otherwise, idle, $\mathbf{Z}$ is the additive white Gaussian noise (AWGN) whose entries are i.i.d circular Gaussian with zero mean and variance $\psi$, i.e., $\mathcal{CN}\left(0,\psi\right)$. Formulating \eqref{eq:1} into more compact expression:
\begin{equation}\label{eq:2}
		\mathbf{Y} = \mathbf{A}\mathbf{X}+\mathbf{Z},
\end{equation}
where matrix $\mathbf{A}$ is the pilot codebook and matrix $\mathbf{X}\in \mathbb{C}^{K\times N_o}$ is a row-sparse matrix where only $K_a$ rows with non-zeros entries are to be detected and estimated, which is a typical compressive sensing model.
\subsection{Channel Model}\label{sec.channel model}
Overall, the channel vector consists of small scale fading coefficients $\mathbf{s}_k$ and a large scale fading coefficient (LSFC) $\varsigma_k$, i.e.,  $\mathbf{h}_k=\sqrt{\varsigma_k}\mathbf{s}_k$. In terms of small scale fading, the geometric model with far-field planar transmission and $L_s$ finite scattering paths is considered. Let $\sigma_{k,l},~\theta_{k,l},~l\in \{1,\ldots,L_s\}$ denote the path strength and angle-of-arrival (AoA) respectively of the $k$-th user at the $l$-th scattering path. The receiving antenna is deemed as a linear array with length of $W=\frac{\lambda_{len}}{2}(M-1)$ where $\frac{\lambda_{len}}{2}$ is half-wavelength and $M$ is a positive constant. In this work, $N_o$ ports are uniformly positioned and the gap width of array elements would be $\frac{W}{N_o-1}$. Thus, the normalized steering response vector corresponding to the $l$-th scattering path of the $k$-th user is \cite{FAS1}:
\begin{equation}\label{eq:3}
	\mathbf{s}_{k,l}= \frac{1}{\sqrt{N_o}}\exp\left(-j\frac{2\pi\left(n-1\right) W}{\left(N_o-1\right)\lambda_{len}}\cos{\theta_{k,l}}\right),n\in\{1,\ldots,N_o\},
\end{equation}
thereby, the small scale fading $\mathbf{s}_k$ can be written as:
\begin{equation}\label{eq:4}
	\mathbf{s}_k = \sum_{l=1}^{L_s}\sigma_{k,l}\mathbf{s}_{k,l}\in \mathbb{C}^{1\times N_o}.
\end{equation}
Regarding LSFC, it is determined by the distance $d_k$ (meters) between the $k$-th user and the BS via a potential large fading model function $\varsigma_k=f(d_k)$, which is normally a very small-valued fraction to be conquered with adequate transmission energy level. For ease of description, small scale fading vector $\mathbf{s}_k$ is normalized with unit entry variance, i.e., $\mathrm{E}\left\{\|\mathbf{s}_k\|_2^2\right\}=N_o$. Thus, $\mathrm{E}\left\{\|\mathbf{h}_k\|_2^2\right\}=N_o\varsigma_k$.
\subsection{Location Model}\label{Location Model}
As depicted in Fig.~\ref{fig:system_model}, for the assumed linear array-based BS, the location of the $k$-th user can be described into polar coordinates $\left(d_k,\theta_k\right)$, where users are randomly located within a service area in a circle sector between radius $d_{ref}\le d_k \le d_{\max}$ and ranging $\theta_{\min}\le\theta_k \le\theta_{\max}$. The distance between the BS and any users will not be smaller that the reference $d_{ref}$, which is either caused by the altitude of BS or a threshold to distinguish the near-field zone. This work only concentrates on the two dimensional location model and assumes far-field planar wave propagation.

Moreover, Notably, scattering path model \eqref{eq:4} can be further categorized by whether there are only non-line of sight (NLOS) components or a mixture of NLOS/LOS components, affecting the strength of $\sigma_{k,l}$. For NLOS-only scattering paths, due to the environment with affluent dispersive obstacles, no signals from direct transmission paths between the user and the BS are observed. For LOS/NLOS scattering, let $K_r$ denote the Rician factor and one of the path strength equals to $\sqrt{\frac{K_r\Omega}{K_r+1}}e^{j\beta_k}$ with $\beta_k$ the arbitrary LOS phase and $\Omega$ a scaled constant and the amplitudes of other path strength are constraint by $\sqrt{\underbrace{\sigma^2_{k,1}+\ldots+\sigma^2_{k,L_s-1}}_{L_s-1 \text{~NLOS components}}}=\sqrt{\frac{\Omega}{K_r+1}}$, i.e., the LOS AoA will have relatively larger path strength than other NLOS AoAs and \textit{the prior PDF of channel coefficients does not necessarily have zero mean due to the offset in path strength}. 
\section{An Explanation on Performance Floor of Greedy Correlation-based Algorithm}\label{inherent_problem}
As illustrated in \cite[Fig.~12, Fig.~13]{FAS1} and \cite[Fig.~2, Fig.~4]{CE5}, a seemingly inherent performance floor phenomenon can be observed in greedy correlation-based algorithms. Specifically, the channel estimation accuracy does not improve with increasing SNR. Instead, the estimation error converges to a floor, which is neither expected nor desirable in practice. In this section, an analytical explanation is provided to reveal rationale behind this phenomenon.

For ease of description, single user transmission is considered and the user index $k$ is omitted. At the $n$-th activated port, the received signal $\mathbf{y}_m$ is given by:
\begin{equation}\label{eq:5}
	\mathbf{y}_n=\mathbf{a}h_n+\mathbf{z}_n,
\end{equation}
where $h_n,n\in\{1,\ldots,N_o\}$ is the $n$-th channel coefficient entry, $\mathbf{a}$ is the pilot, and $\mathbf{z}_n$ is the AWGN at $n$-th port. For greedy correlation-based methods, the detection on active pilot and the estimation of coefficient $h_n$ are generally estimated by:
\begin{equation}\label{eq:6}
	\tilde{y}_n=\frac{\left(\mathbf{a}^\prime\right)^{\mathrm{H}}\mathbf{y}_n}{\left(\mathbf{a}^\prime\right)^{\mathrm{H}}\mathbf{a}}=h_n+\tilde{z}_n,
\end{equation}
where the effective noise follows distribution of $\tilde{z}_n=\frac{\left(\mathbf{a}^\prime\right)^{\mathrm{H}}\mathbf{z}_n}{\left(\mathbf{a}^\prime\right)^{\mathrm{H}}\mathbf{a}}\sim \mathcal{CN}\left(\tilde{z}_n;0,\frac{\psi}{\left(\mathbf{a}^\prime\right)^{\mathrm{H}}\mathbf{a}}\right)$ and $\mathbf{a}^\prime$ is a selected pilot codeword. Therefore, the equalized signal $\tilde{y}_n$ follows distribution of $\mathcal{CN}\left(\tilde{y}_n;\bar{h},\varsigma+\frac{\psi}{\left(\mathbf{a}^\prime\right)^{\mathrm{H}}\mathbf{a}}\right)$, where $\bar{h},\varsigma=f(d)$ are the true mean and variance of channel coefficients.

Treating channel coefficients' PDF priors as unknown parameters (variance $\varsigma^\prime$ and mean $\mathbf{a}^\prime$) to be determined, the activity detection (AD) and channel estimation (CE) problems can be formulated into two maximization problems respectively:
\begin{subequations}\label{eq:7}
\begin{align}
	\text{AD:}&~\hat{\mathbf{a}}=\mathop{\arg\max}_{\mathbf{a}^\prime}~\mathcal{CN}\left(\tilde{y}_n;\bar{h},\varsigma^\prime+\frac{\psi}{\left(\mathbf{a}^\prime\right)^{\mathrm{H}}\mathbf{a}}\right)\label{7-a},\\
	\text{CE:}&~\hat{\varsigma}=\mathop{\arg\max}_{\varsigma^\prime}~\mathcal{CN}\left(\tilde{y}_n;\bar{h},\varsigma^\prime+\frac{\psi}{\left(\hat{\mathbf{a}}\right)^{\mathrm{H}}\mathbf{a}}\right)\label{7-b}.
\end{align}
\end{subequations}
Problem \eqref{7-a} finds the $\mathbf{a}^\prime$ maximize the a posteriori PDF of $\tilde{y}_n$. Clearly, the answer to \eqref{7-a} is $\mathbf{a}^\prime=\mathbf{a}$. Problem \eqref{7-b} aims to find the $\varsigma^\prime$ maximize the a posteriori PDF of $\tilde{y}_n$ with output from \eqref{7-a}. With correct AD, \eqref{7-b} can be given as:
\begin{subequations}\label{eq:8}
	\begin{align}
		\hat{\varsigma}&=\mathop{\arg\max}_{\varsigma^\prime}\frac{1}{\sqrt{2\pi\left(\varsigma^\prime+\psi\right)}}e^{-\frac{|\tilde{y}_n-\bar{h}|^2}{2\left(\varsigma^\prime+\psi\right)}},\\
		&=\mathop{\arg\min}_{\varsigma^\prime}\left( \frac{1}{2}\log2\pi+\log\left(\varsigma^\prime+\psi\right)+\frac{|\tilde{y}_n-\bar{h}|^2}{2\left(\varsigma^\prime+\psi\right)} \right)\label{eq:8-a},\\
		&=|\tilde{y}_n-\bar{h}|^2-\psi\label{eq:8-c},
	\end{align}
\end{subequations}
where \eqref{eq:8-a} is established due to the montonicity of $\log$-function and \eqref{eq:8-c} is calculated by putting the derivative of \eqref{eq:8-c} to zero. We have to note that the output of \eqref{eq:8-c} $\hat{\varsigma}=|\tilde{y}_n-\bar{h}|^2-\psi$ also controls the estimation deviation\footnote{Let $\mathrm{E}\left\{\left(\tilde{y}_n-h_n\right)^2\right\}$ denote estimation deviation. $\mathrm{E}\left\{\left(\tilde{y}_n-h_n\right)^2\right\}=\mathrm{E}\left\{\tilde{y}_n^2\right\}+\mathrm{E}\left\{h_n^2\right\}-2\mathrm{E}\left\{\tilde{y}_n^2-\tilde{z}_n\tilde{y}_n\right\}=\left(\varsigma-\varsigma^\prime\right)-\psi=\varsigma-|\tilde{y}_n-\bar{h}|^2$, indicating how the variance estimation in \eqref{eq:7} and \eqref{eq:8-c} affects the estimation precision.} and is determined by the randomized realization of $\tilde{y}_n$. 

Though, as observed in \eqref{eq:6}, with sufficiently long pilot length and unit pilot power assumption $\mathbf{a}^{\mathrm{H}}\mathbf{a}=1$, the effective noise variance will tends to zero if the SNR goes to infinity $\psi\rightarrow 0$. Yet, it does not guarantee the estimation deviation in \eqref{eq:8-c} also vanishes. Considering $\left(\tilde{y}_n-\bar{h}\right)\sim \mathcal{CN}(\tilde{y}_n-\bar{h};0,\varsigma+\sigma^2_n)$ and \eqref{eq:8-c}, the $\hat{\varsigma}$ can be treated as a chi-squared distribution with 2 degree-of-freedom, mean $\varsigma$ and variance $\left(\varsigma+\psi \right)^2$. Let $\varsigma_n$ and $\hat{\varsigma}_n$ be the true priori variance and the estimated a posteriori variance at the $n$-th port, the total mean square error (MSE) can be calculated as:
\begin{subequations}\label{eq:9}
	\begin{align}
		\mathrm{MSE}&=\frac{1}{N_o}\sum_{n=1}^{N_o}\mathrm{E}_{\tilde{y}_n}\left\{ |\hat{\varsigma}_n-\varsigma_n|^2 \right\},\\
		&=\frac{1}{N_o}\sum_{n=1}^{N_o}\mathrm{var}\left\{\hat{\varsigma_n}\right\}=\left(\varsigma+\psi \right)^2\ge \left(f_{\min}(d)+\psi\right)^2,
	\end{align}
\end{subequations}
which indicates the estimation error does not converge to zero even if the number of activated ports tends to infinity $N_o\rightarrow \infty$ or the SNR tends to infinity $\psi\rightarrow 0$, explaining the performance error floor phenomenon in \cite[Fig.~12, Fig.13]{FAS1}, \cite[Fig.~2, Fig.~4]{CE5}. Notably, the proposed algorithm in this work belongs to the category of self-adaptive Bayesian learning, which distinguishes it from the greedy-based algorithms in \cite{FAS1} and \cite{CE5}. Moreover, the proposed scheme does not suffer from the error floor issue, as the fading variance is explicitly utilized as an additional degree of freedom in the algorithm design. The detailed rationale will be provided in the subsequent sections.
\section{The Proposed Methods: Exploiting Coarse Geographical and Angular Information}\label{proposed}
The starting point of this work stems from the EM-AMP algorithm framework \cite{EM-AMP1}, whose train-of-thought and calculation flows essentially inspire this work. Therefore, in the following, key steps of original EM-AMP are outlined and explained. Yet, applying conventional EM-AMP directly on the FAS model will not solve the inherent problem explained in Section~\ref{inherent_problem}. Following the EM-AMP framework, the contributions of this work are further elaborated. Important symbol notations and their definitions are summarized in Table~\ref{tab:notations}.
\begin{table}[!t]
	\caption{Symbols and Definition}
	\label{tab:notations}
	\centering
	\rowcolors{1}{white}{blue!15}
	\begin{tabular}{cc}
		\toprule 
		\textbf{Notation} & \textbf{Definition} \\ 
		$\mathbf{Y}$ & Received signals \\
		
		$\mathbf{A}$ & Pilot codebook\\
		
		$\mathbf{X}$ & Target signal (row-sparse channel matrix)\\
		
		$\mathbf{Z}$ & Additive while Gaussian noise (AWGN)\\
		
		$\psi$ & AWGN variance\\
		
		$\varsigma_k$ & Large scale fading coefficient (LSFC) of $k$-th user\\
		
		$L_s$ & Number of scattering paths\\
		
		$\sigma_{k,l}$ & Path strength of $k$-th user's $l$-th scattering path\\ 
		
		$\theta_{k,l}$ & Angle-of-arrival (AoA) of $k$-th user at $l$-th scattering path\\
		
		$\lambda_{len}$ & Wavelength\\
		
		$W$ & Antenna length (linear array), $W=\frac{\lambda_{len}\left(M-1\right)}{2}$\\
		
		$\mathbf{s}_{k,l}$ & Small scale fading of $k$-th user at $l$-th scattering path\\
		
		$N_o$ & Active ports num (uniformly distributed)\\
		
		$G$ & Pilot codeword length \\
		
		$d_k$ & Distance between $k$-th user and receiver (2-D domain)\\
		
		$K_r$ & Rician factor\\
		
		$\mu_{k,n}^x$ & Prior Gaussian PDF mean in BG model (target signal)\\
		
		$\phi_{k,n}^x$ & Prior Gaussian PDF variance in BG model (target signal)\\
		
		$\lambda_{k}$ & Activity probability of $k$-th pilot codeword\\
		
		$\mathbf{R}$ & Noise-free output, $\mathbf{R}=\mathbf{AX}$, $r_{g,n}=\mathbf{R}[g,n]$\\
		
		$\hat{\mu}^r_{g,n}$ & Posterior PDF mean of noise-free output\\
		
		$\hat{\phi}^r_{g,n}$ & Posterior PDF variance of noise-free output\\
		
		$\tilde{\mu}^r_{g,n}$ & Posterior expectation of noise-free output\\
		
		$\tilde{\phi}^r_{g,n}$ & Posterior variance of noise-free output\\
		
		$\hat{\mu}^x_{k,n}$ & Posterior PDF mean of target signal\\
		
		$\hat{\phi}^x_{k,n}$ & Posterior PDF variance of target signal\\
		
		$\pi_{k,n}$ & Posterior activity probability of $k$-th pilot at $n$-th port \\
		
		$\tilde{x}_{k,n}$ & Posterior expectation of target signal (channel estimation)\\
		
		$\tilde{\phi}^x_{k,n}$ & Posterior variance of target signal\\
		\bottomrule
	\end{tabular}
\end{table}
\subsection{Retrospects on EM-AMP}
The generalized Bernoulli-Gaussian (BG) mixture model is adopted to describe the priori distribution of $\mathbf{X}$ in \eqref{eq:2}:
\begin{equation}\label{eq:10}
	\begin{aligned}
		&p_{\mathbf{X}}\left(x_{k,n};\lambda_k,\mu^{x}_{k,n},\phi^{x}_{k,n}\right)\\
		&=(1-\lambda_k)\delta\left(x_{k,n}\right)+\lambda_k\mathcal{CN}\left(x_{k,n};\mu^{x}_{k,n},\phi^{x}_{k,n}\right),
	\end{aligned}
\end{equation}
where non-negative constant $\lambda_k$ denotes activity probability of $k$-th codeword, $\mu^{x}_{k,n}$ is the distribution mean of desirable signal component, $\phi^{x}_{k,n}$ is the distribution variance of desirable signal component. The BG mixture model in \eqref{eq:10} is applicable to different array designs or near-field cases since the path responses remain arbitrarily randomized. In addition, when the potential statistical model is complex and requires high fitting accuracy through multiple weighted Gaussian functions, \eqref{eq:10} can be easily adapted. For example, as reported in \cite[Fig.~1]{EM-AMP1}, the GB model provides accurate fitting capability for complex distributions.

 Let $\mathbf{q}_k=\left(\lambda_k,\mu^{x}_{k,n},\phi^{x}_{k,n},\psi\right)$ to aggregate prior parameters to be estimated from noisy observations. In the sequel, the posterior statistics from noisy samples are marked with up-arrow hat, i.e., the estimation of $a$ is $\hat{a}$.

\subsubsection{Computing Posterior Statistics via Priors $\mathbf{q}_k$}Firstly, GAMP models the relationship between the noisy output $y_{g,n}$ and its corresponding noise-free output $r_{g,n}=\mathbf{a}_g^{\mathrm{T}}\mathbf{x}_n$ (noise-free matrix output is $\mathbf{R}$), where $\mathbf{a}_g^{\mathrm{T}}$ is the $g$-th row of $\mathbf{A}$,  $\mathbf{x}_n$ is the $n$-th column of $\mathbf{X}$ and $g\in\left\{1,\ldots,G\right\},n\in\left\{1,\ldots,N_o\right\}$. Therefore, we have conditional PDF of:
\begin{equation}\label{eq:11}
	p_{\mathbf{Y}|\mathbf{R}}(y_{g,n}|r_{g,n};\mathbf{q})=\mathcal{CN}\left(y_{g,n};r_{g,n},\psi\right)
\end{equation}
 With conditional PDF $p_{\mathbf{Y}\mid\mathbf{R}}(y_{g,n}|r_{g,n};\mathbf{q})$, the marginal posterior distribution of noise-free output can be calculated in \eqref{eq:12}
\begin{equation}\label{eq:12}
	\begin{aligned}
			&p_{\mathbf{R}|\mathbf{Y}}(r_{g,n}|\mathbf{y}_n;\hat{\mu}_{g,n}^r,\hat{\phi}_{g,n}^r,\mathbf{q})\\
		&\triangleq\frac{p_{\mathbf{Y}|\mathbf{R}}(y_{g,n}|r_{g,n};\mathbf{q})\mathcal{CN}\left(r_{g,n};\hat{\mu}^r_{g,n},\hat{\phi}_{g,n}^r\right)}{\int_{r} p_{\mathbf{Y}|\mathbf{R}}(y_{g,n}|r;\mathbf{q})\mathcal{CN}\left(r;\hat{\mu}^r_{g,n},\hat{\phi}_{g,n}^r\right)},
	\end{aligned}
\end{equation}
where the denominator is the normalization constant and the estimated quantities $\hat{\mu}_{g,n}^r,\hat{\phi}_{g,n}^r$ vary with iteration $t$ \cite[Table~I, R2-R1]{EM-AMP1} and are calculated by (A1)-(A2) in Algorithm~\ref{alg:algorithm1} respectively. Substituting \eqref{eq:11} into the numerator of \eqref{eq:12} and using identities of $\mathrm{E}\left\{\mathcal{CN}(x;a,A)\mathcal{CN}(x;b,B)\right\} =\frac{aB+bA}{A+B},\mathrm{var}\left\{\mathcal{CN}(x;a,A)\mathcal{CN}(x;b,B)\right\} =\frac{AB}{A+B}$, the posterior statistics of noise-free output are:
\begin{subequations}\label{eq:13}
	\begin{align}
		&\mathrm{E}_{\mathbf{R}|\mathbf{Y}}(r_{g,n}|\mathbf{y}_n;\hat{\mu}_{g,n}^r,\hat{\phi}_{g,n}^r,\mathbf{q})=\frac{\hat{\phi}_{g,n}^r\mathbf{Y}[g,n]+\psi\hat{\mu}_{g,n}^r}{\hat{\phi}_{g,n}^r+\psi},\label{eq:13a}\\
		&\mathrm{var}_{\mathbf{R}|\mathbf{Y}}(r_{g,n}|\mathbf{y}_n;\hat{\mu}_{g,n}^r,\hat{\phi}_{g,n}^r,\mathbf{q})=\frac{\hat{\phi}_{g,n}^{r}\psi}{\hat{\phi}_{g,n}^{r}+\psi},\label{eq:13b}
	\end{align}
\end{subequations}
where we use $\tilde{\mu}^r_{g,n},~\tilde{\phi}^r_{g,n}$ to denote \eqref{eq:13a} and \eqref{eq:13b} respectively.
Subsequently, GAMP approximates the true marginal posterior distribution by:
\begin{equation}\label{eq:14}
	\begin{aligned}
			&p_{\mathbf{X}|\mathbf{Y}}(x_{k,n}|\mathbf{y}_n;\hat{\mu}^{x}_{k,n},\hat{\phi}^{x}_{k,n},\mathbf{q}_k) \\
			&\triangleq \frac{p_\mathbf{x}(x_{k,n};\mathbf{q}_k)\mathcal{CN}(x_{k,n};\hat{\mu}^{x}_{k,n},\hat{\phi}^{x}_{k,n})}{\underbrace{\int_{x} p_\mathbf{x}(x;\mathbf{q}_k)\mathcal{CN}(x;\hat{\mu}^{x}_{k,n},\hat{\phi}^{x}_{k,n})}_{\zeta_{k,n}}},
	\end{aligned}
\end{equation}
where $\hat{\mu}^{x}_{k,n},\hat{\phi}^{x}_{k,n}$ vary by iteration \cite[Table~I, R8-R7]{EM-AMP1} and are calculated by (A7)-(A8) in Algorithm~\ref{alg:algorithm1} respectively and the denominator of \eqref{eq:14} be denoted by:
\begin{subequations}\label{eq:15}
	\begin{align}
		\zeta_{k,n}&=\int_{x} p_\mathbf{x}(x;\mathbf{q}_k)\mathcal{CN}(x;\hat{\mu}^{x}_{k,n},\hat{\phi}^{x}_{k,n})\label{eq:15a}\\
		&=(1-\lambda_k)\mathcal{CN}(0;\hat{\mu}^x_{k,n},\hat{\phi}^x_{k.n})+\label{eq:15b}\\ &\lambda_k\mathcal{CN}(0;\hat{\mu}^x_{k,n}-\mu^x_{k,n},\hat{\phi}^x_{k,n}+\phi^x_{k,n}) \nonumber
	\end{align}
\end{subequations}

Substituting \eqref{eq:10} into \eqref{eq:15a} and using Gaussian-PDF multiplication rule of $\mathcal{CN}(x;a,A)\mathcal{CN}(x;b,B)=\mathcal{CN}(x;\frac{a/A+b/B}{1/A+1/B},\frac{1}{1/A+1/B})\mathcal{CN}(0;a-b,A+B)$, \eqref{eq:15b} can be derived. \textit{The goal is to convert \eqref{eq:14} into a BG structure akin to \eqref{eq:10} and the posterior statistics of target signal can be derived.} Similarly, substituting \eqref{eq:10}, \eqref{eq:15b} into \eqref{eq:14} and using Gaussian-PDF multiplication rule multiple times:
\begin{equation}\label{eq:16}
	\begin{aligned}
		&p_{\mathbf{X}|\mathbf{Y}}(x_{k,n}|\mathbf{y}_n;\hat{\mu}^{x}_{k,n},\hat{\phi}^{x}_{k,n},\mathbf{q}_k)\\
		&\triangleq 
		(1-\pi_{k,n})\delta\left(x_{k,n}\right)+\pi_{k,n}\mathcal{CN}\left(x_{k,n};\gamma_{k,n},\nu_{k,n}\right),
	\end{aligned}
\end{equation}
where parameters including $\pi_{k,n},~\gamma_{k,n}$,~$\nu_{k,n}$ are listed below and for the step-by-step derivations, please refer to the document provided in Sec.~\ref{sec.1-B}:
\begin{subequations}\label{eq:17}
	\begin{align}
		\gamma_{k,n}&\triangleq  \frac{\hat{\mu}^x_{k,n}/\hat{\phi}^x_{k,n}+\mu^x_{k,n}/\phi^x_{k,n}}{1/\hat{\phi}^x_{k,n}+1/\phi^x_{k,n}},\label{eq:17a}\\
		\nu_{k,n}&\triangleq \frac{1}{1/\hat{\phi}^x_{k,n}+1/\phi^x_{k,n}},\label{eq:17b}\\
		\beta_{k,n}&\triangleq \lambda_k\mathcal{CN}\left(\hat{\mu}^x_{k,n};\mu^x_{k,n},\hat{\phi}^x_{k,n}+\phi^x_{k,n}\right),\label{eq:17c}\\
		\pi_{k,n}&\triangleq\frac{1}{1+\left(\frac{\beta_{k,n}}{(1-\lambda_k)\mathcal{CN}(0;\hat{\mu}^x_{k,n},\hat{\phi}^x_{k,n})}\right)^{-1}},\label{eq:17d}
	\end{align}
\end{subequations}
where support probability $0\le\pi_{k,n}\le1$ denotes the likelihood of $x_{k,n}\neq 0$, i.e., the $k$-th pilot codeword at $n$-th antenna. Since the codeword activity is shared among all antennas, the activity likelihood of $k$-th codeword is determined as $\lambda_k=\frac{1}{N_o}\sum_{n=1}^{N_o}\pi_{k,n}$. Based on \eqref{eq:16}, the posterior statistics of target signal are:
\begin{subequations}\label{eq:18}
	\begin{align}
		&\mathrm{E}_{\mathbf{X}|\mathbf{Y}}\left(x_{k,n}|\mathbf{y}_n;\hat{\mu}^x_{k,n},\hat{\phi}^x_{k,n},\mathbf{q}_k\right)=\pi_{k,n}\gamma_{k,n},\label{eq:18a}\\
		&\mathrm{Var}_{\mathbf{X}|\mathbf{Y}}\left(x_{k,n}|\mathbf{y}_n;\hat{\mu}^x_{k,n},\hat{\phi}^x_{k,n},\mathbf{q}_k\right)\label{eq:18b} \\
		&= \pi_{k,n}\left(\nu_{k,n}+|\gamma_{k,n}|^2\right)-|\pi_{k,n}\gamma_{k,n}|^2,\nonumber
	\end{align}
\end{subequations} 
where \eqref{eq:18a} is the estimation on matrix $\mathbf{X}$ and in the sequel, we use $\tilde{x}_{k,n},~\tilde{\phi}^x_{k,n}$ to denote \eqref{eq:18a} and \eqref{eq:18b}.
\subsubsection{Learning $\mathbf{q}_k$ From Noisy Observations}
Observing from \eqref{eq:11} to \eqref{eq:17d}, the calculation flows of GAMP require all parameters in $\mathbf{q}_k$ as input, which will be learned during the iteration. Interestingly, the estimation procedures based on the posterior statistics from GAMP have accomplished the \textit{E-step} (the rational behind EM can be referred at \cite[Eq.18-Eq.21]{EM-AMP1}). The \textit{M-step} is then formulated as:
\begin{equation}\label{eq:19}
	\mathbf{q}_k^{t+1}=\arg\max_{\mathbf{q}_k^t}\hat{\mathrm{E}}\{\ln p_{\mathbf{X}}(\mathbf{X};\mathbf{q}_k)\mid\mathbf{Y};\mathbf{q}_k^t\},
\end{equation}
where $\hat{\mathrm{E}}$ represents the use of GAMP's posterior approximation and and annotation $\left(t\right)$ and $\left(t+1\right)$ denote the parameter at current and next iteration respectively. \textit{One feasible interpretation on \eqref{eq:19} is that the $\ln p(\mathbf{X};\mathbf{q}_k)$ functions as likelihood function by prior distribution $p_{\mathbf{X}}(\mathbf{X};\mathbf{q}_k)$ and the averaging by posterior approximation, i.e., $\hat{\mathrm{E}}\left\{\cdot\right\}$, plays as weight function.} An example of expression expansion of \eqref{eq:19} can be referred in \eqref{eq:21}. The prior parameters in $\mathbf{q}_k$ follow the update rules listed as follows and for the step-by-step derivations, please refer to the document provided in Sec.~\ref{sec.1-B}:
\begin{subequations}\label{eq:20}
	\begin{align}
		&\lambda_k^{t+1} \triangleq \frac{1}{K}\sum_{n=1}^{N_o}\pi_{k,n},\label{eq:20a}\\
		&\psi^{t+1} \triangleq \frac{1}{GK}\sum_{k=1}^{K}\sum_{n=1}^{N_o}\left(|y_{g,n}-\tilde{\mu}^r_{g,n}|^2+\tilde{\phi}^r_{g,n}\right),\label{eq:20b}\\
		&\mu^{x,t+1}_{k,n} \triangleq \frac{\sum_{k=1}^{K}\pi_{k,n}\gamma_{k,n}}{\lambda_k^{t+1}K},\label{eq:20c}\\
		&\phi^{x,t+1}_{k,n}\triangleq\frac{1}{\lambda_k^{t+1}K}\sum_{k=1}^{K}\pi_{k,n}\left(|\mu^{x,t}_{k,n}-\gamma_{k,n}|^2+\nu_{k,n}\right)\label{eq:20d},
	\end{align}
\end{subequations}
where $\tilde{\mu}^r_{g,n},~\tilde{\phi}^r_{g,n}$ refer to \eqref{eq:13a} and \eqref{eq:13b} respectively, $\gamma_{k,n}$ is calculated by \eqref{eq:17a}, and $\nu_{k,n}$ is calculated by \eqref{eq:17b}. 
\begin{algorithm}[t!] 
	\small
	\caption{Algorithm Baseline I, EM-AMP for FAS}
	\label{alg:algorithm1}
	\KwIn{$\mathbf{Y}$, $\mathbf{A}$, $K$, $N_o$, $G$,  $\psi$, $T_{\max}$}
	
	\textbf{Initialize:}\\
	
	$\forall k:\lambda_k^1=\frac{G}{K}\max_{a>0}\frac{1-\frac{2K}{G}[(1+a^2)\Phi(-a)-a\mathcal{N}(a;0,1)]}{1+a^2-2[(1+a^2)\Phi(-a)-a\mathcal{N}(a;0,1)]}$\hfill \text{(I1)} \\
	
	$\forall k,n:\phi^{x,1}_{k,n}=\frac{\sum_{g=1}^{G}\left|\mathbf{Y}[g,n]\right|^{2}-M\sigma_{n}^{2}}{\sum_{g=1}^{G}\sum_{k=1}^{K}\lvert \mathbf{A}[g,k]\rvert^{2}\lambda_{k}^1}$, $\mu^{x,1}_{k,n}=0$\hfill \text{(I2)} \\
	
	$\forall k,n:\tilde{x}^1_{k,n}=\int_{x}^{\infty}xp_{\mathbf{X}}\left(x;\lambda^1_k,\mu^{x,1}_{k,n},\phi^{x,1}_{k,n}\right)\mathrm{d}x$\hfill \text{(I3)} \\
	
	$\forall k,n:\tilde{\phi}^1_{k,n}=\int_{x}^{\infty}\lvert x-\tilde{x}^1_{k,n}\rvert^2 p_{\mathbf{X}}\left(x;\lambda^1_k,\mu^{x,1}_{k,n},\phi^{x,1}_{k,n}\right)\mathrm{d}x$\hfill \text{(I4)} \\
	
	$\forall g,n:\hat{s}^0_{g,n}=0$\hfill \text{(I5)}
	
	\ForEach{$t={1,2},\cdots,T_{\max}$}{
		AMP part:
		
		$\forall g,n: \hat{\phi}^{r,t}_{g,n}=\sum_{k=1}^{K}\lvert \mathbf{A}[g,k]\rvert^2 \phi^{x,t}_{k,n}$\hfill \text{(A1)}
		
		$\forall g,n: \hat{\mu}^{r,t}_{g,n}=\sum_{k=1}^{K}\mathbf{A}[g,k]\tilde{x}^t_{k,n}-\hat{\phi}^{r,t}_{g,n}\hat{s}^{t-1}_{g,n}$\hfill \text{(A2)}
		
		$\forall g,n: \tilde{\phi}^{r,t}_{g,n}=\frac{\hat{\phi}_{g,n}^{r,t}\mathbf{Y}[g,n]+\psi\hat{\mu}_{g,n}^{r,t}}{\hat{\phi}_{g,n}^{r,t}+\psi}$\hfill \text{(A3)}
		
		$\forall g,n: \tilde{\mu}^{r,t}_{g,n}=\frac{\hat{\phi}_{g,n}^{r,t}\psi}{\hat{\phi}_{g,n}^{r,t}+\psi}$\hfill \text{(A4)}
		
		$\forall g,n: \hat{\phi}^{s,t}_{g,n}=\frac{\hat{\phi}^{r,t}_{g,n}-\tilde{\phi}^{r,t}_{g,n}}{\left(\hat{\phi}^{r,t}_{g,n}\right)^2}$\hfill \text{(A5)}
		
		$\forall g,n:
		\hat{s}^t_{g,n}=\frac{\tilde{\mu}^{r,t}_{g,n}-\hat{\mu}^{r,t}_{g,n}}{\hat{\phi}^{r,t}_{g,n}}$\hfill \text{(A6)}
		
		$\forall k,n:
		\hat{\phi}^{x,t}_{k,n}=\left(\sum_{g=1}^{G}\lvert \mathbf{A}[g,k] \rvert^2\hat{\phi}^{s,t}_{g,n} \right)^{-1}$\hfill \text{(A7)}
		
		$\forall k,n:
		\hat{\mu}^{x,t}_{k,n}=\tilde{x}^t_{k,n}+\hat{\phi}^{x,t}_{k,n}\sum_{g=1}^{G}\left(\mathbf{A}[g,k]\right)^*\hat{s}^t_{g,n}$\hfill \text{(A8)}
		
		$\forall k,n:
		\gamma_{k,n}\triangleq  \frac{\hat{\mu}^{x,t}_{k,n}/\hat{\phi}^{x,t}_{k,n}+\mu^{x,t}_{k,n}/\phi^{x,t}_{k,n}}{1/\hat{\phi}^{x,t}_{k,n}+1/\phi^{x,t}_{k,n}}$\hfill \text{(B1)}
		
		$\forall k,n:
		\nu_{k,n}\triangleq \frac{1}{1/\hat{\phi}^{x,t}_{k,n}+1/\phi^{x,t}_{k,n}}$\hfill \text{(B2)}
		
		$\forall k,n:
		\beta_{k,n}\triangleq \lambda^t_k\mathcal{CN}\left(\hat{\mu}^{x,t}_{k,n};\mu^{x,t}_{k,n},\hat{\phi}^{x,t}_{k,n}+\phi^{x,t}_{k,n}\right)$\hfill \text{(B3)}
		
		$\forall k,n:
		\pi_{k,n}\triangleq\frac{1}{1+\left(\frac{\beta_{k,n}}{(1-\lambda^t_k)\mathcal{CN}(0;\hat{\mu}^{x,t}_{k,n},\hat{\phi}^{x,t}_{k,n})}\right)^{-1}}$\hfill \text{(B4)}
		
		$\forall k,n:
		\tilde{\phi}^{t+1}_{k,n}=\pi_{k,n}\left(\nu_{k,n}+|\gamma_{k,n}|^2\right)-|\pi_{k,n}\gamma_{k,n}|^2$\hfill \text{(A9)}
		
		$\forall k,n:
		\tilde{x}^{t+1}_{k,n}=\pi_{k,n}\gamma_{k,n}$\hfill \text{(A10)}
		
		EM part:
		
		$\forall k: \lambda_k^{t+1} \triangleq \frac{1}{K}\sum_{n=1}^{N_o}\pi_{k,n}$\hfill \text{(E1)}
		
		$\forall k,n: \mu^{x,t+1}_{k,n} \triangleq \frac{\sum_{k=1}^{K}\pi_{k,n}\gamma_{k,n}}{\lambda_k^{t+1}K}$\hfill \text{(E2)}
		
		$\forall k,n: \phi^{x,t+1}_{k,n}\triangleq\frac{\sum_{k=1}^{K}\pi_{k,n}\left(|\mu^{x,t}_{k,n}-\gamma_{k,n}|^2+\nu_{k,n}\right)}{\lambda_k^{t+1}K}$\hfill \text{(E3)}
	}
	\KwOut{AD liklihood $\lambda_{k},k\in\left\{1,\ldots,K\right\}$ and effective CE $\tilde{\mathbf{X}}[k,n]=\tilde{x}^{t+1}_{k,n},k\in\left\{1,\ldots,K\right\}, n\in\left\{1,\ldots,N_o\right\}$.}
\end{algorithm}

In the sequel, the conventional EM-AMP on model \eqref{eq:2} is deemed as a benchmark and the calculation flow is summarized in Algorithm~\ref{alg:algorithm1}, where (A3) and (A4) are derived in \eqref{eq:13b} and \eqref{eq:13a}, (A5) and (A6) are intermediate parameters \cite[Table I, R5-R6]{EM-AMP1}, (B1)-(B4) are derived from \eqref{eq:17a} to \eqref{eq:17d}, (A9)-(A10) are derived in \eqref{eq:18a} and \eqref{eq:18b} and (E1)-(E3) are derived from \eqref{eq:20a} to \eqref{eq:20d}. Meanwhile, since the estimation on AWGN variance is not in the scope of this work, the noise variance $\psi$ is treated as a known parameter for simplicity to reduce the implementation redundancy without loss of generality. The initialization (I1)-(I5) follows the principles in \cite[Section~II-D]{EM-AMP1}.  
\begin{algorithm}[t!] 
	\small
	\caption{Proposed I, Exploiting Coarse Geographical Information}
	\label{alg:algorithm2}
	\KwIn{$\mathbf{Y}$, $\mathbf{A}$, $K$, $N_o$, $G$,  $\psi$, $T_{\max}$, $\phi^x_{\min}$, $\phi^x_{\max}$}
	
	\textbf{Initialize:}~(I1)-(I5) of Algorithm~\ref{alg:algorithm1}
	
	\ForEach{$t={1,2},\cdots,T_{\max}$}{
		AMP part: (A1)-(A10) of Algorithm~\ref{alg:algorithm1}
		
		EM part:
		
		(E1) of Algorithm~\ref{alg:algorithm1}
		
		$\forall k: \phi^{x,t+1}_{k,n}=\left(\tilde{x}^{t+1}_{k,n}-\mu^{x,t}_{k,n}\right)^2-\tilde{\phi}^{t+1}_{k,n}$\hfill \text{(P1)}	
		
		\uIf{$\phi^{x,t+1}_{k,n}>\phi^{x}_{\max}$}{$\phi^{x,t+1}_{k,n}=\phi^{x}_{\max}$}\ElseIf{$\phi^{x,t+1}_{k,n}<\phi^{x}_{\min}$}{$\phi^{x,t+1}_{k,n}=\phi^{x}_{\min}$}
		
		(E2) of Algorithm~\ref{alg:algorithm1}
	}
	\KwOut{AD liklihood $\lambda_{k},k\in\left\{1,\ldots,K\right\}$ and effective CE $\tilde{\mathbf{X}}[k,n]=\tilde{x}^{t+1}_{k,n},k\in\left\{1,\ldots,K\right\}, n\in\left\{1,\ldots,N_o\right\}$.}
\end{algorithm}
\subsection{Proposed I: Exploiting Coarse Geographical Information}
Inspired by the observations in \eqref{eq:9}, the information on geographical information of single user plays essential role in terms of estimation accuracy. In our case, the geographical information refers to the LSFC $\varsigma_k$ of channel model in Section~\ref{sec.channel model}, which directly determines the variance $\phi^x_{k,n}$ of prior distribution $p_{\mathbf{X}}\left(x_{k,n};\lambda_k,\mu^{x}_{k,n},\phi^{x}_{k,n}\right)$ in \eqref{eq:10}. More importantly, since the distance information on 2-D plane and the variance $\phi^x_{k,n}$ are assumed to be correlated by $\phi^x_{k,n}=f(d_k)$ \cite{LSFC1,LSFC2,LSFC3}, one could exploit this additional geographical information to update $\phi^x_{k,n}$. In the sequel, let $\phi^x_{k,n}(d_k)$ represent prior variance is a function of distance.

Following the EM principle and incremental updating rule \cite{EM-AMP1,EM1,EM2}, the distance information can be sequentially updated and estimated akin to \eqref{eq:19}. Due to the independence among users' location, the update rule for $d_k$ can be formulated by:
\begin{equation}\label{eq:21}
	\begin{aligned}
		&d_k^{t+1} =\mathop{\arg\max}_{d_{ref}\le d_k\le d_{\max}}\sum_{n=1}^{N_o} \hat{\mathrm{E}}\left\{\ln p_{\mathbf{X}}\left(x_{k,n};\mathbf{q}_k\right)|\mathbf{Y},\mathbf{q}^t_k\right\}\\
		&=\mathop{\arg\max}_{d_{ref}\le d_k\le d_{\max}}\sum_{n=1}^{N_o}\underbrace{\int_{x_{k,n}}p_{\mathbf{X}|\mathbf{Y}}(x_{k,n}|\mathbf{y}_n;\mathbf{q}^t_k)\ln p_{\mathbf{X}}\left(x_{k,n};\mathbf{q}^t_k\right)}_{\triangleq J\left(\phi_{k,n}\right)},
	\end{aligned}
\end{equation}
where posterior PDF $p_{\mathbf{X}|\mathbf{Y}}(x_{k,n}|\mathbf{y}_n;\mathbf{q}^t_k)$ and prior PDF $p_{\mathbf{X}}\left(x_{k,n};\mathbf{q}^t_k\right)$ are identical to \eqref{eq:16} and \eqref{eq:10} respectively and we denote the integral in \eqref{eq:21} by $J\left(\phi_{k,n}\right)$ in the sequel.

Meanwhile, the integral area should be split into separate domains considering that the logarithmic term in $J\left(\phi^x_{k,n}\right)$ has different expressions:
\begin{equation}\label{eq:22}
	p_{\mathbf{X}}\left(x_{k,n};\mathbf{q}^t_k\right)=\left\{\begin{matrix}
		(1-\lambda^t_k)\delta\left(x_{k,n}\right),& x_{k,n}=0\\
		\lambda^t_k\mathcal{CN}\left(x_{k,n};\mu^{x,t}_{k,n},\phi^{x,t}_{k,n}\left(d_k\right)\right),& x_{k,n}\neq 0.
	\end{matrix}\right.
\end{equation}
\begin{figure*}[t!]
	\normalsize
	\begin{subequations}\label{eq:23}
		\begin{align}
			&J\left(\phi^x_{k,n}\right) = \lim_{\epsilon \to 0}\int_{x_{k,n}\in \mathcal{B}_{\epsilon}} p_{\mathbf{X}|\mathbf{Y}}(x_{k,n}|\mathbf{y}_n;\mathbf{q}^t_k)\ln p_{\mathbf{X}}\left(x_{k,n};\mathbf{q}^t_k\right) 		
			+		
			\lim_{\epsilon \to 0}\int_{x_{k,n}\in \overline{\mathcal{B}}_{\epsilon}} p_{\mathbf{X}|\mathbf{Y}}(x_{k,n}|\mathbf{y}_n;\mathbf{q}^t_k)
			\ln p_{\mathbf{X}}\left(x_{k,n};\mathbf{q}^t_k\right) \nonumber\\
			&= \underbrace{\lim_{\epsilon \to 0}\int_{x_{k,n}\in \mathcal{B}_{\epsilon}} p_{\mathbf{X}|\mathbf{Y}}(x_{k,n}|\mathbf{y}_n;\mathbf{q}^t_k)
				\ln \left[(1-\lambda^t_k)\delta\left(x_{k,n}\right)\right]}_{\triangleq C_{k,n}} 
			+	
			\lim_{\epsilon \to 0}\int_{x_{k,n}\in \overline{\mathcal{B}}_{\epsilon}} p_{\mathbf{X}|\mathbf{Y}}(x_{k,n}|\mathbf{y}_n;\mathbf{q}^t_k) 
			\ln \left[\lambda^t_k\mathcal{CN}\left(x_{k,n};\mu^{x,t}_{k,n},\phi^{x,t}_{k,n}\left(d_k\right)\right)\right] \label{eq:23a}\\
			&=C_{k,n}
			+
			\lim_{\epsilon \to 0}\int_{x_{k,n}\in \overline{\mathcal{B}}_{\epsilon}} p_{\mathbf{X}|\mathbf{Y}}(x_{k,n}|\mathbf{y}_n;\mathbf{q}^t_k) 
			\ln \left[
			\frac{\lambda^t_k}{\pi \phi^{x,t}_{k,n}\left(d_k\right)}\exp\left\{-\frac{\lvert x_{k,n}-\mu^{x,t}_{k,n}\rvert^2}{\phi^{x,t}_{k,n}\left(d_k\right)}
			\right\}
			\right]\nonumber\\
			&=C_{k,n}
			+\ln \left(
			\frac{\lambda^t_k}{\pi \phi^{x,t}_{k,n}\left(d_k\right)}\right)
			\underbrace{ \lim_{\epsilon \to 0}\int_{x_{k,n}\in \overline{\mathcal{B}}_{\epsilon}} p_{\mathbf{X}|\mathbf{Y}}(x_{k,n}|\mathbf{y}_n;\mathbf{q}^t_k)}_{\triangleq \pi_{k,n}}
			-\frac{1}{\phi^{x,t}_{k,n}\left(d_k\right)}\underbrace{\lim_{\epsilon \to 0}\int_{x_{k,n}\in \overline{\mathcal{B}}_{\epsilon}} p_{\mathbf{X}|\mathbf{Y}}(x_{k,n}|\mathbf{y}_n;\mathbf{q}^t_k) \lvert x_{k,n}-\mu^{x,t}_{k,n}\rvert^2}_{\triangleq V_{k,n}}\label{eq:23b}\\
			&=C_{k,n}+\pi_{k,n}\ln \left(
			\frac{\lambda^t_k}{\pi \phi^{x,t}_{k,n}\left(d_k\right)}\right)-\frac{V_{k,n}}{\phi^{x,t}_{k,n}\left(d_k\right)}.\label{eq:23c}
		\end{align}
	\end{subequations}
		\hrulefill
\end{figure*}
Accordingly, the integral area is split into two parts denoted by $\mathcal{B}_{\epsilon}=\left[-\epsilon,\epsilon\right]$ and $\overline{\mathcal{B}}_{\epsilon}=\mathbb{C}\setminus \mathcal{B}_{\epsilon}$, where $\epsilon\to 0$ controls the borders between $\mathcal{B}_{\epsilon}$ and $\overline{\mathcal{B}}_{\epsilon}$. The integral process is give in \eqref{eq:23} with result of $J\left(\phi^x_{k,n}\right)=C_{k,n}+\pi_{k,n}\ln \left(
\frac{\lambda^t_k}{\pi \phi^{x,t}_{k,n}\left(d_k\right)}\right)-\frac{V_{k,n}}{\phi^{x,t}_{k,n}\left(d_k\right)}$. In \eqref{eq:23a}, $C_{k,n}$ is shown as a constant irrelevant to $\phi^x_{k,n}$. For \eqref{eq:23b}, two major integral components ($ \pi_{k,n}$ and $V_{k,n}$) are calculated as:
\begin{subequations}\label{eq:24}
	\begin{align}
		&\lim_{\epsilon \to 0}\int_{x_{k,n}\in \overline{\mathcal{B}}_{\epsilon}} p_{\mathbf{X}|\mathbf{Y}}(x_{k,n}|\mathbf{y}_n;\mathbf{q}^t_k) \nonumber\\
		&=\lim_{\epsilon \to 0}\int_{x_{k,n}\in \overline{\mathcal{B}}_{\epsilon}}\pi_{k,n}\mathcal{CN}\left(x_{k,n};\gamma_{k,n},\nu_{k,n}\right)=\pi_{k,n},\label{eq:24a}\\
		&V_{k,n}=\lim_{\epsilon \to 0}\int_{x_{k,n}\in \overline{\mathcal{B}}_{\epsilon}} p_{\mathbf{X}|\mathbf{Y}}(x_{k,n}|\mathbf{y}_n;\mathbf{q}^t_k) \lvert x_{k,n}-\mu^{x,t}_{k,n}\rvert^2 \nonumber \\
		&=\left[\mathrm{E}_{\mathbf{X}|\mathbf{Y}}\left(x_{k,n}|\mathbf{y}_n;\mathbf{q}^t_k\right)-\mu^{x,t}_{k,n}\right]^2-\mathrm{Var}_{\mathbf{X}|\mathbf{Y}}\left(x_{k,n}|\mathbf{y}_n;\mathbf{q}^t_k\right)\nonumber \\
		&=\left(\underbrace{\pi_{k,n}\gamma_{k,n}}_{\tilde{x}^{t+1}_{k,n}}-\mu^{x,t}_{k,n}\right)^2-\underbrace{\pi_{k,n}\left(\nu_{k,n}+|\gamma_{k,n}|^2\right)+|\pi_{k,n}\gamma_{k,n}|^2}_{\tilde{\phi}^{t+1}_{k,n}},\label{eq:24b}
	\end{align}
\end{subequations}
where $\mathrm{E}_{\mathbf{X}|\mathbf{Y}}\left(x_{k,n}|\mathbf{y}_n;\mathbf{q}^t_k\right)$ and $\mathrm{Var}_{\mathbf{X}|\mathbf{Y}}\left(x_{k,n}|\mathbf{y}_n;\mathbf{q}^t_k\right)$ are identical to the statistics in \eqref{eq:18} omitting irrelevant terms and have been calculated before EM update during (A9)-(A10) in Algorithm~\ref{alg:algorithm1}. Therefore, the EM maximization expression in \eqref{eq:21} is converted into:
\begin{equation}\label{eq:25}
	\small
	\begin{aligned}
		&d_k^{t+1}=f^{-1}\left(\phi^{x,t+1}_{k,n}\right),~\phi^{x}_{\min}\le \phi^{x,t}_{k,n}\le \phi^{x}_{\max}\\
		&=\mathop{\arg\max}_{ \phi^{x,t}_{k,n}}\sum_{n=1}^{N_o}C_{k,n}+\pi_{k,n}\ln \left(
		\frac{\lambda^t_k}{\pi \phi^{x,t}_{k,n}\left(d_k\right)}\right)-\frac{V_{k,n}}{\phi^{x,t}_{k,n}\left(d_k\right)}\\
		&\Rightarrow \mathop{\arg\max}_{ \phi^{x,t}_{k,n}}\sum_{n=1}^{N_o}
		\pi_{k,n}\ln \left(
		\frac{\lambda^t_k}{\pi \phi^{x,t}_{k,n}\left(d_k\right)}\right)-\frac{V_{k,n}}{\phi^{x,t}_{k,n}\left(d_k\right)}\\
		&=\mathop{\arg\min}_{\phi^{x,t}_{k,n}}\sum_{n=1}^{N_o}
		\pi_{k,n}\ln \left(
		\frac{\pi}{\lambda^t_k}\right)+\pi_{k,n}\ln \left(\phi^{x,t}_{k,n}\left(d_k\right)\right)+\frac{V_{k,n}}{\phi^{x,t}_{k,n}\left(d_k\right)}\\
		&\Rightarrow \mathop{\arg\min}_{\phi^{x}_{\min}\le \phi^{x,t}_{k,n}\le \phi^{x}_{\max}}\sum_{n=1}^{N_o}\pi_{k,n}\ln \left(\phi^{x,t}_{k,n}\left(d_k\right)\right)+\frac{V_{k,n}}{\phi^{x,t}_{k,n}\left(d_k\right)},
	\end{aligned}
\end{equation}
where components $C_{k,n}$ and $\pi_{k,n}\ln \left(\phi^{x,t}_{k,n}\left(d_k\right)\right)$ are omitted since they are irrelevant to $\phi^{x,t}_{k,n}$. Since the variance contributed by LSFC should be identical among all receiving antennas, one can put the first-derivative of \eqref{eq:25} to zero and find a closed-form solution to update the prior PDF variance:
\begin{equation}\label{eq:26}
\phi^{x,t+1}_{k,n}=\left\{\begin{matrix}
	\phi^x_{\min},& \mathrm{if}~\frac{\sum_{n=1}^{N_o}V_{k,n}}{\sum_{n=1}^{N_o}\pi_{k,n}}<\phi^x_{\min} \\ 
	\frac{\sum_{n=1}^{N_o}V_{k,n}}{\sum_{n=1}^{N_o}\pi_{k,n}}, & \mathrm{if}~\phi^x_{\min}\le\frac{\sum_{n=1}^{N_o}V_{k,n}}{\sum_{n=1}^{N_o}\pi_{k,n}}\le \phi^x_{\max}\\
	\phi^x_{\max},& \mathrm{if}~\frac{\sum_{n=1}^{N_o}V_{k,n}}{\sum_{n=1}^{N_o}\pi_{k,n}}>\phi^x_{\max}
\end{matrix}\right.
\end{equation}
where intermediate parameter $V_{k,n}$ is calculated in \eqref{eq:24b}. Hereby, we summarize the EM-AMP exploiting geographical feature for FAS in Algorithm~\ref{alg:algorithm2} where (P1) is derived in \eqref{eq:24b} and \eqref{eq:26}. Chronologically, one can observe the working flow of signal variance $\phi^{x}_{k,n}$:
\begin{itemize}
	\item[-] The prior of target signal variance $\phi^{x}_{k,n}$ first participates in the calculation of noise-free output statistics in (A1)-(A3).
	\item[-] Then, it affects intermediate residuals in (A5)-(A6) and thus influences empirical statistics from noisy observations by (A7)-(A8).
	\item[-] Finally, the posterior PDF statistics are calculated in (B1)-(B4) relevant to AD and CE.
\end{itemize}
\subsection{Proposed II: Exploiting Angular Information}
Inspired by the mechanism how prior PDF variance $\phi^{x}_{k,n}$ affects the calculation flow of EM-AMP, we also observe that \textit{the value of effective CE $\tilde{x}_{k,n}$ can also propagate through similar update flows and eventually affect the final results, i.e., $\tilde{x}_{k,n}\rightarrow\text{(A2)}~\hat{\mu}^r_{g,n}\rightarrow \text{(A3)}~\tilde{\phi}^r_{g,n}\rightarrow \text{(A5)}~\hat{\phi}^{s}_{g,n}\rightarrow \text{(A7)}~\hat{\phi}^x_{k,n}\rightarrow \text{(A8)}~\hat{\mu}^x_{k,n}\rightarrow\text{(B1)-(B4)}$}. Therefore, this part exploits the inherent angular diversity of target $\tilde{x}_{k,n}$ and then offer insights into potential methods to refine the effective estimation.

Considering channel row-vector $\mathbf{s}_k$ and rewriting \eqref{eq:4} column-wisely:
\begin{equation}\label{eq:27}
	\begin{aligned}
		\mathbf{s}^{\mathrm{T}}_k 
		&= \sum_{l=1}^{L_s}\sigma_{k,l}\mathbf{s}^{\mathrm{T}}_{k,l}
		=\left[\mathbf{s}^{\mathrm{T}}_{k,1},\ldots,\mathbf{s}^{\mathrm{T}}_{k,L_s}\right]
		\cdot
		\begin{bmatrix}
			\sigma_{k,1}\\
			\vdots \\
			\sigma_{k,L_s}
		\end{bmatrix} \\
		&=\underbrace{ \left[\mathbf{w}_1,\mathbf{w}_2,\ldots,\mathbf{w}_{N_s}\right]}_{\text{Steering Response Codebook $\mathbf{W}$}}\cdot \underbrace{\begin{bmatrix}
			0\\
			\sigma_{k,1}\\
			\vdots\\
			\sigma_{k,l}\\
			\vdots\\
			0
		\end{bmatrix}}_{\text{$L_s$-sparse vector $\tilde{\boldsymbol{\sigma}}_k$}},
	\end{aligned}
\end{equation}
where $\mathbf{w}_i,i\in\{1,\ldots,N_s\}$ is the steering response vectors generated with AoA samples $\theta_i$ by \eqref{eq:3}, $N_s$ is the number of AoA samples, the non-zero elements in the $L_s$-sparse vector $\tilde{\boldsymbol{\sigma}}_k$ denotes the activity of corresponding columns in $\mathbf{W}$ and \textit{the value of none-zero elements corresponds to the path strength scaled by LSFC.} Therefore, among many potential methods, one feasible way to exploit the sparsity feature in angular domain is treating \eqref{eq:27} as a \textit{sparse linear regression model} and conduct compressive sensing method to express effective estimation $\tilde{x}_{k,n}$ within the span of steering responses in $\mathbf{W}$, i.e., the effective CE at each round of EM-AMP can be refined by $\tilde{\mathbf{x}}_k^{\mathrm{T}}\leftarrow \mathbf{W}\tilde{\boldsymbol{\sigma}}_k$ where $\tilde{\mathbf{x}}_k^{\mathrm{T}}$ is the $k$-th row of the target estimation matrix and $\tilde{\boldsymbol{\sigma}}_k$ is calculated by solving the problem in \eqref{eq:27} based on estimation $\tilde{\mathbf{x}}_k^{\mathrm{T}}$. 
\begin{algorithm}[t!]
	\SetAlgoLined
	\KwIn{Received Signal $\mathbf{Y}$, Codebook $\mathbf{C}$, Activity $\mathcal{S}$}
	\tcp{This algorithm will automatically be reduced to OMP if receiver has single antenna.}
	\KwOut{Active Indices and Estimated Channel Vectors}
	Initialization: $k\leftarrow 0$, $\mathbf{R}\leftarrow \mathbf{Y}$, $\mathcal{S}\leftarrow \emptyset$, $\mathbf{H}\leftarrow \mathbf{0}$ \;
	\While{$k \le K_a$}{
		$i_{k}\leftarrow \mathop{\arg\max}\limits_{i_k}\|\mathbf{R}^{\mathrm{H}}\mathbf{C}[:,i_k]\|_2/\|\mathbf{C}[:,i_k]\|_2$\;
		$\mathcal{S}\leftarrow \mathcal{S} \cup i_{k}$\;
		Projection Span $\mathbf{\Phi}\leftarrow \mathbf{C}\left[:,\{\mathcal{S}\}\right]$ \;
		$\mathbf{R}\leftarrow \left(\mathbf{R}-\mathbf{\Phi}\mathbf{\Phi}^{\dagger}\mathbf{Y}\right )$ and $\left(\cdot\right)^{\dagger}$ denotes Moore-Penrose inverse\;
		$k\leftarrow k+1$\;}
	Active Indices $\mathcal{S}=\{i_{1},i_{2},\ldots,i_{K_a}\}$ and 
	Estimation $\left \{ \mathbf{h}_1^{\mathrm{T}}, \mathbf{h}_2^{\mathrm{T}},\ldots,\mathbf{h}_{K_a}^{\mathrm{T}} \right \}\leftarrow \mathbf{C}\left[:,\{\mathcal{S}\}\right]^{\dagger}\mathbf{Y}.$
	\caption{Algorithm Baseline II, SOMP}
	\label{alg:SOMP_MIMO}
\end{algorithm}
\begin{algorithm}[t!] 
	\small
	\caption{Proposed II, Exploiting Angular Information}
	\label{alg:algorithm4}
	\KwIn{$\mathbf{Y}$, $\mathbf{A}$, $K$, $N_o$, $G$,  $\psi$, $T_{\max}$}
	
	\textbf{Initialize:}~(I1)-(I5) of Algorithm~\ref{alg:algorithm1}
	
	\ForEach{$t={1,2},\cdots,T_{\max}$}{
		AMP part: (A1)-(A10) of Algorithm~\ref{alg:algorithm1}
		
		EM part: (E1)-(E3) of Algorithm~\ref{alg:algorithm1}
		
		$\forall k, \tilde{\mathbf{x}}_k^{\mathrm{T}}=\tilde{\mathbf{X}}[k,:]$, Refine $\tilde{\mathbf{x}}_k^{\mathrm{T}}\approx \mathbf{W}\tilde{\boldsymbol{\sigma}}_k$ by model \eqref{eq:27}, which express the estimation within span of steering responses in codebook $\mathbf{W}$ \hfill \text{(P2)} 
	}
	\KwOut{AD liklihood $\lambda_{k},k\in\left\{1,\ldots,K\right\}$ and effective CE $\tilde{\mathbf{X}}[k,n]=\tilde{x}^{t+1}_{k,n},k\in\left\{1,\ldots,K\right\}, n\in\left\{1,\ldots,N_o\right\}$.}
\end{algorithm}

For ease of equipment, we adopt matching pursuit method to make refinement on the effective CE. Firstly, we introduce another baseline algorithm named as simultaneous orthogonal matching pursuit (SOMP) \cite{ODMA_iotj}. The calculation flows are summarized in Algorithm~\ref{alg:SOMP_MIMO}. SOMP identifies the codeword in $\mathbf{C}$ with the highest correlation to the signal residual matrix $\mathbf{R}$. The index of this codeword is added to the support set $\mathcal{S}$. Subsequently, the projection span matrix $\mathbf{\Phi}$ is formed using the detected codewords, and the residual is updated as described in step 6 of Algorithm~\ref{alg:SOMP_MIMO}. Using the identified active codewords, the corresponding channel coefficients are then estimated. 

Reminding that SOMP can be readily adopted to solve the AD and CE problem by model \eqref{eq:2} and will automatically be reduced to OMP if the receiver has only one antenna. Compared with OMP, SOMP can utilize the multiple measurement vector (MMV) feature \cite{SOMP} akin to step (E1) in Algorithm~\ref{alg:algorithm1}. Hereby, we summarize the EM-AMP exploiting angular information for FAS in Algorithm~\ref{alg:algorithm4} where (P2) constrains the effective CE within the span of steering responses and $\tilde{\boldsymbol{\sigma}}_{k}$ is the estimated path strength. Notably, there are various ways to detect and estimate the potential AoAs such as EM-based \cite{EM_AoA}, DFT-based \cite{CE5}, subspace-search kind \cite{FAS-ISAC1}, etc.

 Furthermore, we will demonstrate that lower estimation error will be generated if angular information is considered. Without loss of generality, assuming perfect detection and considering single user case for CE with signal model $\mathbf{Y}=\mathbf{a}\mathbf{h}+\mathbf{Z}$ with LS solution to CE of $\tilde{\mathbf{h}}= \left(\mathbf{a}^{\mathrm{H}}\mathbf{a}\right)^{-1}\mathbf{a}^{\mathrm{H}}\mathbf{Y}$, the MSE of CE without angular information $M$ is calculated as:
 \begin{equation}\label{eq:add-28}
 	\begin{aligned}
 		&M = \mathrm{E}\left\{\left(\mathbf{h}-\tilde{\mathbf{h}}\right)\left(\mathbf{h}-\tilde{\mathbf{h}}\right)^{\mathrm{H}} \right\}\\
 		&\xrightarrow[]{\text{by Eq.~\ref{eq:add-30}}} \mathrm{E}\left\{ \left(\mathbf{a}^{\mathrm{H}}\mathbf{a}\right)^{-1}\mathbf{a}^{\mathrm{H}}\left(\mathbf{Z}\mathbf{Z}^{\mathrm{H}}\right)\mathbf{a}\left(\mathbf{a}^{\mathrm{H}}\mathbf{a}\right)^{-1} \right\}.
 	\end{aligned}
 \end{equation} 
 \begin{figure*}[htp]
 \normalsize
 \begin{subequations}\label{eq:add-30}
 	\begin{align}
 		M &= \mathrm{E}\left\{\left(\mathbf{h}-\tilde{\mathbf{h}}\right)\left(\mathbf{h}-\tilde{\mathbf{h}}\right)^{\mathrm{H}} \right\}
 		= \mathrm{E}\left\{\left(\mathbf{h}-\left(\mathbf{a}^{\mathrm{H}}\mathbf{a}\right)^{-1}\mathbf{a}^{\mathrm{H}}\mathbf{Y}\right)\left(\mathbf{h}-\left(\mathbf{a}^{\mathrm{H}}\mathbf{a}\right)^{-1}\mathbf{a}^{\mathrm{H}}\mathbf{Y}\right)^{\mathrm{H}} \right\} \\
 		& = \mathrm{E}\left\{\left(\mathbf{h}-\left(\mathbf{a}^{\mathrm{H}}\mathbf{a}\right)^{-1}\mathbf{a}^{\mathrm{H}}\left(\mathbf{a}\mathbf{h}+\mathbf{Z}\right) \right)\left(\mathbf{h}-\left(\mathbf{a}^{\mathrm{H}}\mathbf{a}\right)^{-1}\mathbf{a}^{\mathrm{H}}\left(\mathbf{a}\mathbf{h}+\mathbf{Z}\right)\right)^{\mathrm{T}} \right\} \\		
 		&=
 		\mathrm{E}\begin{Bmatrix}
 			\bigg( \mathbf{h}\mathbf{h}^{\mathrm{H}}-
 			\mathbf{h}\left(\mathbf{a}\mathbf{h}+\mathbf{Z}\right)^{\mathrm{H}}\mathbf{a}\left(\mathbf{a}^{\mathrm{H}}\mathbf{a}\right)^{-1}-
 			\left(\mathbf{a}^{\mathrm{H}}\mathbf{a}\right)^{-1}\mathbf{a}^{\mathrm{H}}\left(\mathbf{a}\mathbf{h}+\mathbf{Z}\right)\mathbf{h}^{\mathrm{H}}
 			\\
 			+ \left(\mathbf{a}^{\mathrm{H}}\mathbf{a}\right)^{-1}\mathbf{a}^{\mathrm{H}}\left(\mathbf{a}\mathbf{h}+\mathbf{Z}\right)\left(\mathbf{a}\mathbf{h}+\mathbf{Z}\right)^{\mathrm{H}}\mathbf{a}\left(\mathbf{a}^{\mathrm{H}}\mathbf{a}\right)^{-1}\bigg)
 		\end{Bmatrix}\\
 		&\approx \mathrm{E}\left\{ \left(\mathbf{a}^{\mathrm{H}}\mathbf{a}\right)^{-1}\mathbf{a}^{\mathrm{H}}\left(\mathbf{Z}\mathbf{Z}^{\mathrm{H}}\right)\mathbf{a}\left(\mathbf{a}^{\mathrm{H}}\mathbf{a}\right)^{-1} \right\}.
 	\end{align}
 		\hrulefill
 \end{subequations}
 \end{figure*}
 \begin{figure*}[htp]
 \normalsize
 \begin{subequations}\label{eq:add-31}
 	\begin{align}
 		M_{\mathrm{angular}}&=
 		\mathrm{E}\left\{\left(\boldsymbol{\sigma}-\tilde{\boldsymbol{\sigma}}\right)\mathbf{U}\mathbf{U}^{\mathrm{H}}\left(\boldsymbol{\sigma}-\tilde{\boldsymbol{\sigma}}\right)^{\mathrm{H}}\right\}\\
 		&= \mathrm{E}\left\{\left(\boldsymbol{\sigma}-\mathbf{a}^{\mathrm{H}}\mathbf{Y}\mathbf{U}^{\mathrm{H}}\left(\mathbf{U}\mathbf{U}^{\mathrm{H}}\right)^{-1}\right)\mathbf{U}\mathbf{U}^{\mathrm{H}}\left(\boldsymbol{\sigma}-\mathbf{a}^{\mathrm{H}}\mathbf{Y}\mathbf{U}^{\mathrm{H}}\left(\mathbf{U}\mathbf{U}^{\mathrm{H}}\right)^{-1}\right)^{\mathrm{H}}\right\} \\
 		& = \mathrm{E}\left\{\left(\boldsymbol{\sigma}-\mathbf{a}^{\mathrm{H}}\left(\mathbf{a}\left(\boldsymbol{\sigma}\mathbf{U}\right)+\mathbf{Z}\right)\mathbf{U}^{\mathrm{H}}\left(\mathbf{U}\mathbf{U}^{\mathrm{H}}\right)^{-1}\right)\mathbf{U}\mathbf{U}^{\mathrm{H}}\left(\boldsymbol{\sigma}-\mathbf{a}^{\mathrm{H}}\left(\mathbf{a}\left(\boldsymbol{\sigma}\mathbf{U}\right)+\mathbf{Z}\right)\mathbf{U}^{\mathrm{H}}\left(\mathbf{U}\mathbf{U}^{\mathrm{H}}\right)^{-1}\right)^{\mathrm{H}}\right\}\\		
 		&=\mathrm{E}\begin{Bmatrix}
 			\bigg( \boldsymbol{\sigma}\mathbf{U}\mathbf{U}^{\mathrm{H}}\boldsymbol{\sigma}^{\mathrm{H}}-\boldsymbol{\sigma}\mathbf{U}\mathbf{U}^{\mathrm{H}}\left(\mathbf{U}\mathbf{U}^{\mathrm{H}}\right)^{-1}\mathbf{U}\left(\mathbf{a}\boldsymbol{\sigma}\mathbf{U}+\mathbf{Z}\right)^{\mathrm{H}}\mathbf{a}-\mathbf{a}^{\mathrm{H}}\left(\mathbf{a}\boldsymbol{\sigma}\mathbf{U}+\mathbf{Z}\right)\mathbf{U}^{\mathrm{H}}\left(\mathbf{U}\mathbf{U}^{\mathrm{H}}\right)^{-1}\mathbf{U}\mathbf{U}^{\mathrm{H}}\boldsymbol{\sigma}^{\mathrm{H}}
 			\\
 			+ \mathbf{a}^{\mathrm{H}}\left(\mathbf{a}\boldsymbol{\sigma}\mathbf{U}+\mathbf{Z}\right)\mathbf{U}^{\mathrm{H}}\left(\mathbf{U}\mathbf{U}^{\mathrm{H}}\right)^{-1}\mathbf{U}\mathbf{U}^{\mathrm{H}}\left(\mathbf{U}\mathbf{U}^{\mathrm{H}}\right)^{-1}\mathbf{U}\left(\mathbf{a}\boldsymbol{\sigma}\mathbf{U}+\mathbf{Z}\right)^{\mathrm{H}}\mathbf{a}\bigg)
 		\end{Bmatrix}\\
 		&\approx\mathrm{E}\left\{\mathbf{a}^{\mathrm{H}}\mathbf{Z}\mathbf{U}^{\mathrm{H}}\left(\mathbf{U}\mathbf{U}^{\mathrm{H}}\right)^{-1}\mathbf{U}\mathbf{U}^{\mathrm{H}}\left(\mathbf{U}\mathbf{U}^{\mathrm{H}}\right)^{-1}\mathbf{U}\mathbf{Z}^{\mathrm{H}}\mathbf{a} \right\}.
 	\end{align}
 \end{subequations}
 	\hrulefill
 \end{figure*}
  \begin{figure}[!t]
 	\centering
 	\includegraphics[width=\columnwidth]{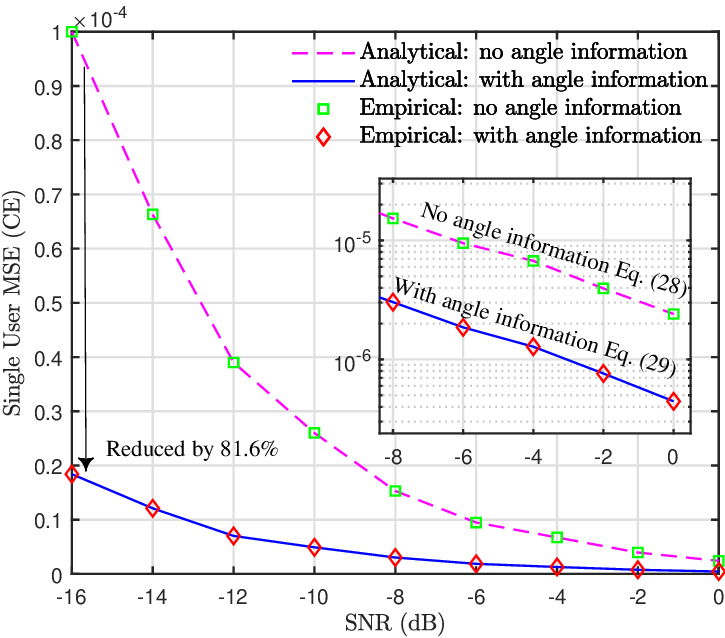}
 	\caption{Illustration of analytical/empirical user MSE (CE) under different SNR (dB) with antenna length constant $M=64$, $N_o=16$ active ports, $G=200$ pilot length and other parameters in Table~\ref{tab:system configurations}. Perfect priors (AD and AoAs are known) are assumed to generate analytical results by \eqref{eq:add-28}, \eqref{eq:add-29} and empirical results are generated by $\tilde{\mathbf{h}}= \left(\mathbf{a}^{\mathrm{H}}\mathbf{a}\right)^{-1}\mathbf{a}^{\mathrm{H}}\mathbf{Y}$ and $\tilde{\boldsymbol{\sigma}}=\mathbf{a}^{\mathrm{H}}\mathbf{Y}\mathbf{U}^{\mathrm{H}}\left(\mathbf{U}\mathbf{U}^{\mathrm{H}}\right)^{-1}$ respectively for no and with angular information.}\label{sim:MSE_comparison}
 \end{figure}
 By assuming independence among signal (target signal, background noise and pilot codeword) components, one can obtain $M = \mathrm{E}\left\{ \left(\mathbf{a}^{\mathrm{H}}\mathbf{a}\right)^{-1}\mathbf{a}^{\mathrm{H}}\left(\mathbf{Z}\mathbf{Z}^{\mathrm{H}}\right)\mathbf{a}\left(\mathbf{a}^{\mathrm{H}}\mathbf{a}\right)^{-1} \right\}$ following the derivations in \eqref{eq:add-30}. Subsequently, we investigate the MSE $M_{\mathrm{angular}}$ with angular information. Similarly, without loss of generality, considering single user case for CE with signal model with angular information $\mathbf{Y}=\mathbf{a}\mathbf{h}+\mathbf{Z}=\mathbf{a}\left(\boldsymbol{\sigma}\mathbf{U}\right)+\mathbf{Z}$, where the rows of $\mathbf{U}$ are the independent steering response vectors generated in \eqref{eq:3} assumed to be known. With known $\mathbf{a}$ and some maneuvers of transpose and shift, one can have the corresponding LS solution is $\tilde{\boldsymbol{\sigma}}=\mathbf{a}^{\mathrm{H}}\mathbf{Y}\mathbf{U}^{\mathrm{H}}\left(\mathbf{U}\mathbf{U}^{\mathrm{H}}\right)^{-1}$. Thereby, the MSE of CE with angular information is calculated as:
 \begin{equation}\label{eq:add-29}
 	\begin{aligned}
 		&M_{\mathrm{angular}}
 		=\mathrm{E}\left\{\left(\boldsymbol{\sigma}-\tilde{\boldsymbol{\sigma}}\right)\mathbf{U}\mathbf{U}^{\mathrm{H}}\left(\boldsymbol{\sigma}-\tilde{\boldsymbol{\sigma}}\right)^{\mathrm{H}}\right\}\\
 		&\xrightarrow[]{\text{by Eq.~\ref{eq:add-31}}} \mathrm{E}\left\{\mathbf{a}^{\mathrm{H}}\mathbf{Z}\mathbf{U}^{\mathrm{H}}\left(\mathbf{U}\mathbf{U}^{\mathrm{H}}\right)^{-1}\mathbf{U}\mathbf{U}^{\mathrm{H}}\left(\mathbf{U}\mathbf{U}^{\mathrm{H}}\right)^{-1}\mathbf{U}\mathbf{Z}^{\mathrm{H}}\mathbf{a} \right\}.
 	\end{aligned}
 \end{equation}
 
 By assuming independence among signal (target signal, background noise, pilot codeword and response steering vectors) components, one can obtain $M_{\mathrm{angular}} = \mathrm{E}\left\{\mathbf{a}^{\mathrm{H}}\mathbf{Z}\mathbf{U}^{\mathrm{H}}\left(\mathbf{U}\mathbf{U}^{\mathrm{H}}\right)^{-1}\mathbf{U}\mathbf{U}^{\mathrm{H}}\left(\mathbf{U}\mathbf{U}^{\mathrm{H}}\right)^{-1}\mathbf{U}\mathbf{Z}^{\mathrm{H}}\mathbf{a} \right\}$ following the derivations in \eqref{eq:add-31}.

However, due to the randomness of $\mathbf{U}$, tracing its distribution is not feasible. We illustrate the differences between \eqref{eq:add-28} and \eqref{eq:add-29} in Fig~\ref{sim:MSE_comparison} listing explicit parameter configurations. Notably, leveraging angular information significantly reduces CE derivations, thereby substantially enhancing CE precision. The consistency between the analytical and empirical results in Fig.~\ref{sim:MSE_comparison} further validates the derived MSE results in \eqref{eq:add-28} and \eqref{eq:add-29}. \textit{In summary, utilizing angular information for FAS represents a promising research direction.}

\subsection{Complexity Analyses}
In this section, the computational complexities of relevant algorithms are analyzed:
\begin{itemize}
	\item[1)] 
	For \textbf{baseline Algorithm~\ref{alg:algorithm1}}, the computational complexity of the conventional EM-AMP is primarily driven by steps (A1)--(A2) and (A7)--(A8), where matrix multiplications occur, with a complexity of $ \mathcal{O}(4KGN_o) $ per iteration. For the EM component, updating the prior activity likelihood $ \lambda_k $ requires $ \mathcal{O}(KN_o) $ per iteration, while updating the prior PDF variance and mean incurs $ \mathcal{O}(2KN_o) $ per iteration. Overall, the total computational complexity is approximately $ \mathcal{O}(4KGN_o + 3KN_o) $, which is irrelevant to $K_a$ and thus favorable for massive connectivity.
	\item[2)] For \textbf{the proposed Algorithm~\ref{alg:algorithm2}} exploiting geographical feature, it incurs slightly additional complexity in updating the prior variance, which still is marginal relative to the computations in the AMP part.
	\item[3)] 
	For \textbf{baseline Algorithm~\ref{alg:SOMP_MIMO}}, the computational complexity of SOMP is primarily driven by the Moore-Penrose inverse at each iteration, with a complexity of $ \mathcal{O}(G^3) $. Given that SOMP performs identical procedures $ K_a $ times, the overall complexity scales as $ \mathcal{O}(N_o K_a G^3) $.
	
	\item[4)] 
	
	For \textbf{the proposed Algorithm~\ref{alg:algorithm4}}, the computational complexity of exploiting angular information is approximately $ \mathcal{O}(K N_o^3) $ when using refinement solutions via OMP. However, it must be emphasized that many alternative methods exist for channel refinement, many of which exhibit low complexity characteristics such as EM-based \cite{EM_AoA}, DFT-based \cite{CE5}, etc.
\end{itemize} 

The AMP categories are better suited for massive connectivity, as their computational complexity is fixed to the codebook size $ K $ and independent of the number of active users $ K_a $. In contrast, the complexity of greedy-based algorithms, such as SOMP, increases with $ K_a $. Consequently, for scenarios with low activity, greedy-based algorithms incur lower complexity overhead, whereas AMP categories offer significantly reduced computational complexity in cases with a large number of active users.

\begin{table}[t!]
	\caption{System Configurations}
	\label{tab:system configurations}
		\centering
	\begin{tabular}{lll}
		\hline
		Parameter       & Definitions               & Setups      \\ \hline
		$d_{\max}$      & Far field upper range     & 100 meters  \\
		$d_{def}$       & Far field lower range     & 20 meters   \\
		$\theta_{\max}$ & FAS AoA angle upper range & 150 degrees \\
		$\theta_{\min}$ & FAS AoA angle upper range & 30 degrees  \\
		$f(d_k)$        & LSFC function             & $d_k^{-2}$  \\
		$L_s$           & Scattering path num       & 3           \\ 
		$T_{\max}$      & EM-AMP iteration upper range & 50\\
		$K$ 			& Total user num			& 1000\\
		$N_s$           & AoA sample num			& 121 (resolution $1^o$) \\
		\hline
	\end{tabular}
\end{table}
\section{Numerical Results}\label{Numerical Results}
In this part, numerical results under various setups are illustrated to verify the proposed algorithms. System configurations are listed in Table~\ref{tab:system configurations} and remain unchanged if not particularly stated. Moreover, the LSFC model i.e., $f(d_k)=d_k^{-2}$, has been adopted in many influential works \cite{LSFC1,LSFC2,LSFC3}. The performance metrics \cite{AMP2,AMP3}include activity detection error (ADE), channel estimation normalized mean square error (NMSE), mean square error (MSE) of prior PDF variance estimation:
\begin{subequations}\label{eq:32}
	\small
	\begin{align}
	\mathrm{ADE}&= 1-\frac{\lvert \mathcal{A}\cap \tilde{\mathcal{A}} \rvert}{K_a}, \\
	\mathrm{NMSE}&= \frac{\mathrm{E}\left\{\|\mathbf{h}_k-\tilde{\mathbf{h}}_k\|_2^2\right\}}{\mathrm{E}\left\{\|\mathbf{h}_k\|_2^2\right\}}, \\
	\mathrm{MSE}&= \mathrm{E}\left\{\lvert \phi^x_k-\tilde{\phi}^x_k \rvert^2\right\},
	\end{align}
\end{subequations}
where $\mathcal{A},\mathbf{h}_k,\phi^x_k$ denote the true activity set, channel coefficients prior PDF variance of the $k$-th active user, and $\tilde{\mathcal{A}},\tilde{\mathbf{h}}_k,\tilde{\phi}^x_k$ are corresponding estimation. Some crucial notes are listed: 
\begin{itemize}
	\item[-] For AD, SOMP requires prior of activity and outputs only $K_a$ active candidates while others do not need prior of activity.
	\item[-] For equity (greedy-based algorithms require known $K_a$) and for simple ADE calculation, we verdict codewords as active with largest $K_a$ likelihood $\lambda_{k}$ in AMP-based algorithms.
	\item[-] Only the estimation of correctly detected users are averaged for NMSE (CE) and MSE (prior PDF variance).
\end{itemize}
 Moreover, the \textit{received} SNR is defined by:
\begin{equation}\label{eq:29}
	\mathrm{SNR}=\frac{\|\mathbf{a}_k\|_2^2\mathrm{E}\left\{\|\mathbf{h}_k\|_2^2\right\}}{\mathrm{E}\left\{\|\mathbf{Z}\|_{\mathsf{F}}^2\right\}}=\frac{N_o\bar{\varsigma}_k}{\psi GN_o}=\frac{\bar{\varsigma}_k}{G\psi},
\end{equation}
where $\|\mathbf{a}_k\|_2^2=1$ since this work has assumed pilot with unit power from the beginning in Section~\ref{sec.channel model} and $\bar{\varsigma}_k=\frac{1}{K_a}\sum_{k=1}^{K_a}\varsigma_k$ is the averaged LSFC among all active users thus the value of AWGN variance need to adapt to each round of realization due to randomness of user location. Moreover, the algorithm benchmarks are conventional EM-AMP \cite{EM-AMP1} in Algorithm~\ref{alg:algorithm1}, SOMP+LS \cite{CE5} in Algorithm~\ref{alg:SOMP_MIMO} and AoA codebook-based method \cite{FAS_channel 0.2}.
\begin{figure}[!t]
	\centering
	\includegraphics[width=\columnwidth]{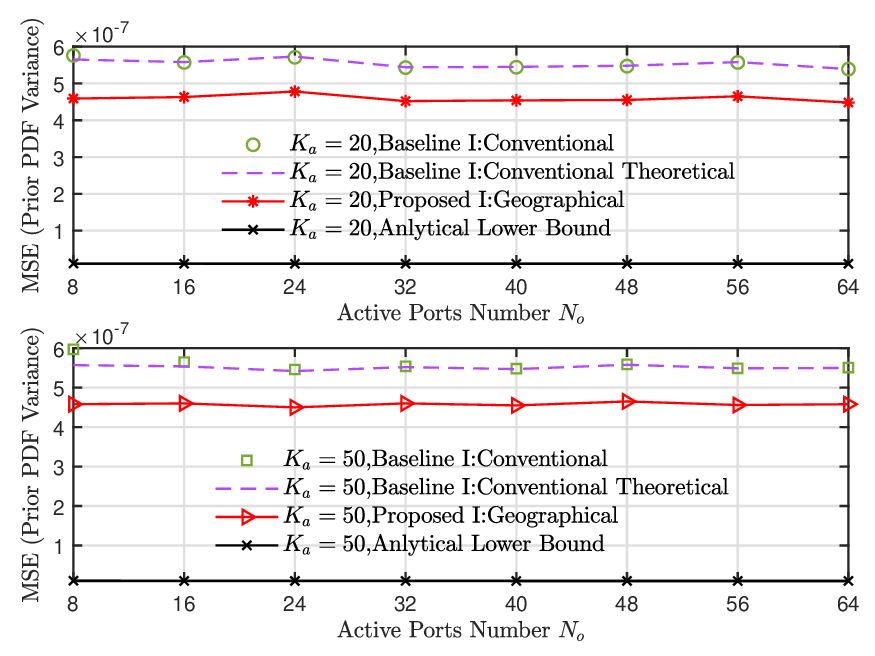}
	\caption{Verification on analytical results in \eqref{eq:9} under different active ports number $N_o$ with antenna length constant $M=64$, pilot length $G=200$, different active users $K_a\in\{20,50\}$ and $\mathrm{SNR}=-10$ dB. Rician factors are $K_r=0$ (NLOS). Performance baselines in comparison are conventional EM-AMP, theoretical performance of conventional EM-AMP and analytical lower bound in \eqref{eq:9}.}\label{sim:variance_MSE}
\end{figure}
\subsubsection{Verification on Theoretical Analyses of \eqref{eq:9}}
In Section~\ref{inherent_problem}, we derived the analytical MSE for prior PDF variance in \eqref{eq:9}, i.e., with the aforementioned system configurations $\mathrm{MSE}=\underbrace{\left(\phi^x+\psi\right)}_{\text{Theoretical}}\ge \underbrace{(d_{\max}^{-2}+\psi)^2}_{\text{Lower Bound}}$. To verify the correctness of the analytical result, Fig.~\ref{sim:variance_MSE} illustrates the analytical and empirical MSE versus different number of active ports $N_o$ of different EM-AMP, the conventional and the proposed algorithm exploiting geographical features summarized in Algorithm~\ref{alg:algorithm2}. Other parameter setups include antenna length constant $M=64$, pilot length $G=200$, different active users $K_a\in\{20,50\}$, $\mathrm{SNR}=-10$ dB and Rician factors are $K_r=0$ (NLOS).

In Fig.~\ref{sim:variance_MSE}, for both 20 and 50 active users, the derived theoretical results align closely with those of the conventional EM-AMP, confirming the accuracy of the theoretical analyses in \eqref{eq:9}. This alignment arises because the conventional EM-AMP approach similarly generates estimations by maximizing the posterior PDF, as described in \eqref{eq:7}. Furthermore, the proposed Algorithm-I, which leverages geographical information, exhibits lower MSE compared to the conventional method, fundamentally demonstrating its superior estimation precision. This improvement will be further validated in the subsequent sections.
\begin{figure}[!t]
	\centering
	\includegraphics[width=\columnwidth]{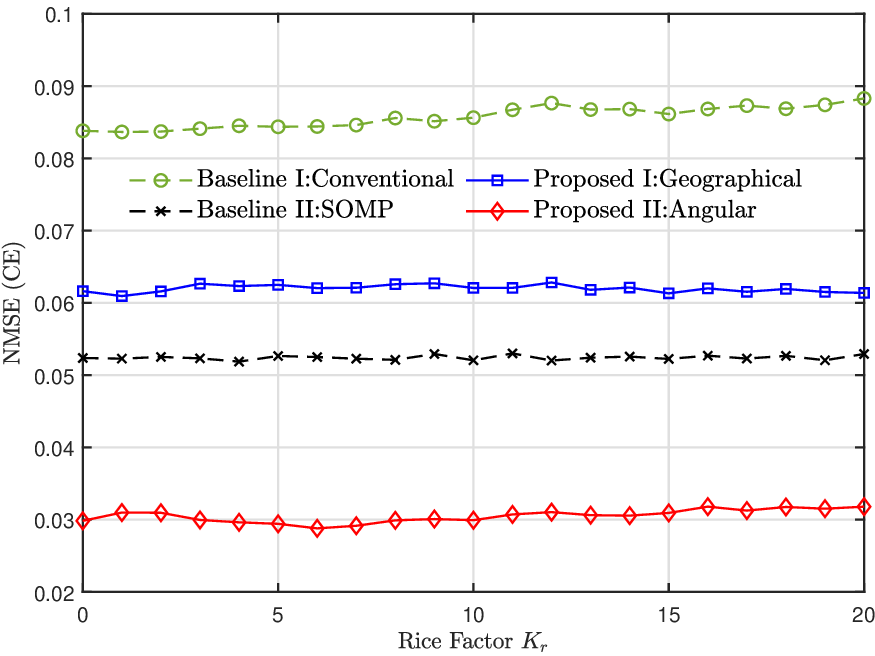}
	\caption{Illustration of NMSE (CE) under different Rician factor $K_r$ with $M=64$ antenna length constant, $N_o=16$ active ports, $K_a=10$ active users and $\mathrm{SNR}=-10$ dB. Performance baselines in comparison are conventional EM-AMP and SOMP.}\label{sim:rice_factor}
\end{figure}
\subsubsection{Performance versus Rician Factor}

In Section~\ref{Location Model}, it is noted that the LOS component of the channel model may introduce an offset in the prior distribution PDF, potentially impacting the performance of AMP due to its dependence on accurate prior distribution modeling. Therefore, in Fig.~\ref{sim:rice_factor}, the NMSE (CE) performance is evaluated under varying Rician factors $ K_r $, with a fixed antenna length $ M=64 $, $ N_o=16 $ active ports, $ K_a=10 $ active users, and $ \mathrm{SNR}=-10 $ dB. 

As the Rician factor increases, the channel model transitions from purely NLOS to a combination of LOS and NLOS components. However, no significant fluctuations are observed across all algorithms, except for a slight increase in the NMSE of the conventional EM-AMP when comparing $ K_r=20 $ to $ K_r=0 $.

\begin{figure}[htp]
	\centering
	\includegraphics[width=3 in]{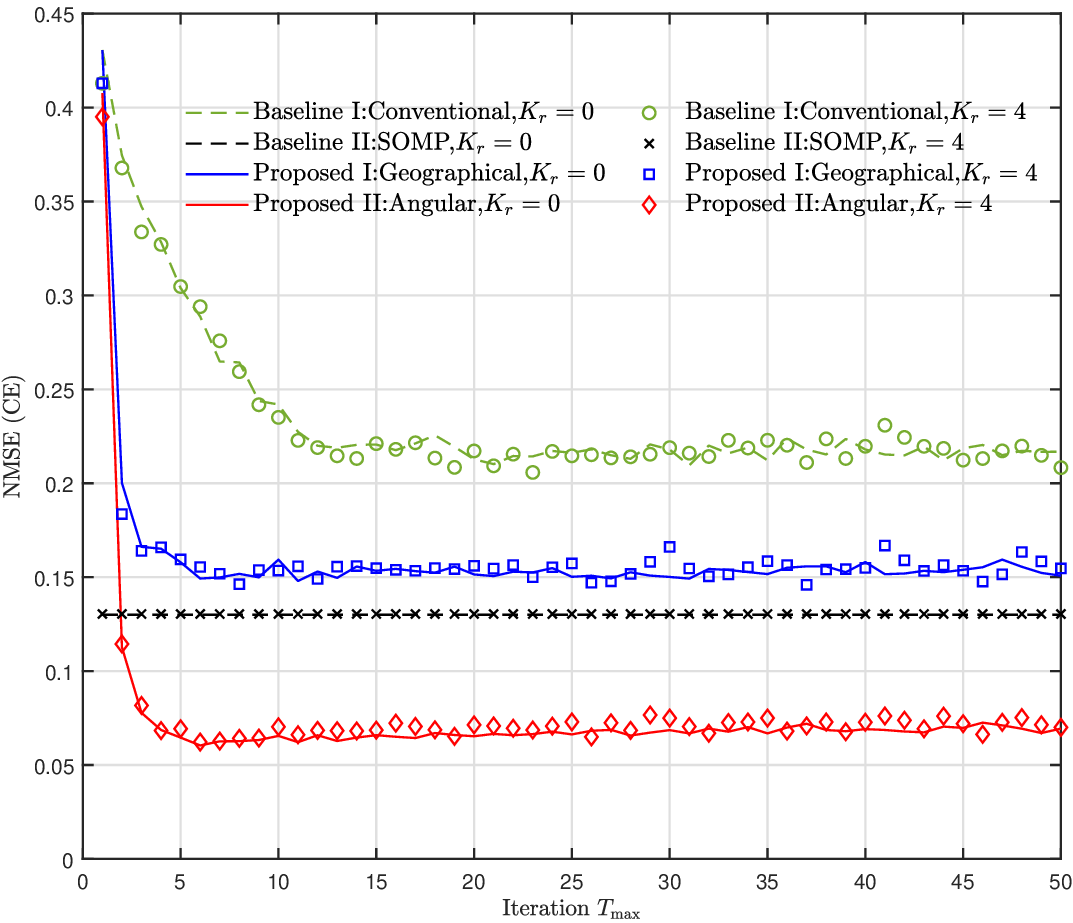}
	\caption{Illustration of convergence behavior of different algorithms with antenna length constant $M=64$, active ports num $N_o=16$, pilot length $G=200$, $K_a=10$ active users and $\mathrm{SNR}=-14$ dB. Rician factors are $K_r=0$ (NLOS) and $K_r=4$ (LOS/NLOS). Performance baselines in comparison are conventional EM-AMP and SOMP.}\label{sim:convergence_CE_nmse}
\end{figure}
\subsubsection{Convergence Behavior}
Unlike greedy-based algorithms, EM-AMP requires sufficient iteration rounds to converge the estimation results, with convergence behavior significantly impacting computational complexity. Thus, it is crucial to examine the convergence behavior of the proposed algorithms. In Fig.~\ref{sim:convergence_CE_nmse}, the convergence behavior of various algorithms is depicted with a fixed antenna length $ M=64 $, $ N_o=16 $ active ports, pilot length $ G=200 $, $ K_a=10 $ active users, and $ \mathrm{SNR}=-14 $ dB, under Rician factors $ K_r=0 $ (NLOS) and $ K_r=4 $ (LOS/NLOS). 

Notably, the Rician factor has negligible impact on performance, consistent with prior findings. Furthermore, the proposed Algorithms I and II demonstrate superior convergence speed, requiring approximately 5 iterations to converge, compared to nearly 15 iterations for the conventional EM-AMP, indicating a triple faster convergence speed. Additionally, the proposed Algorithm I achieves a 25\% lower NMSE than the conventional method, while Algorithm II attains an NMSE close to 0.05, the best among all algorithms, improving CE precision by approximately 75\% compared to the conventional approach. Intuitively, the rapid convergence can be attributed to a more precise search for key parameters within the constrained estimation span, facilitated by valuable signal priors derived from geographical and angular information.
\begin{figure}[!t]
	\centering
	\includegraphics[width=\columnwidth]{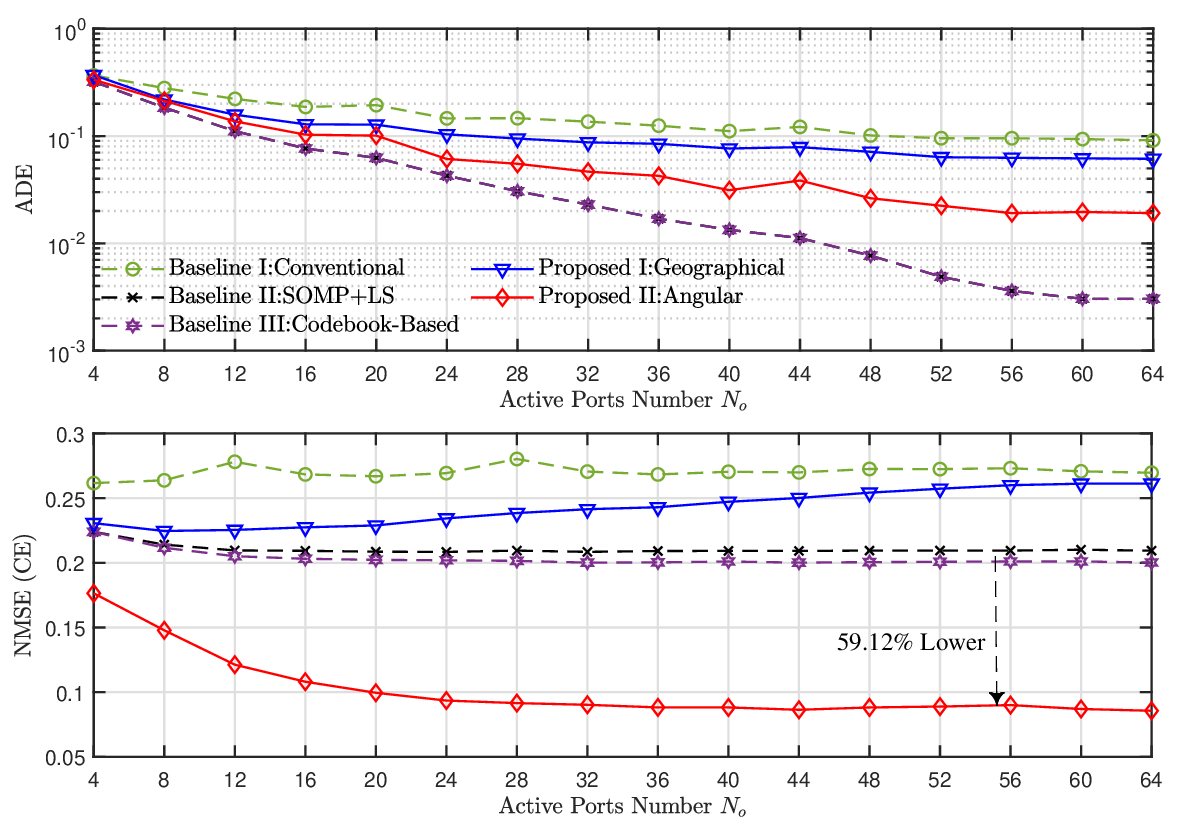}
	\caption{Illustration of ADE (Up) and CE NMSE (Down) of different algorithms versus active ports number $N_o$ with antenna length constant $M=64$, pilot length $G=200$, $K_a=50$ active users and $\mathrm{SNR}=-15$ dB. Rician factors are $K_r=0$ (NLOS). Performance baselines in comparison are conventional EM-AMP \cite{EM-AMP1}, SOMP+LS \cite{CE5} and AoA codebook-based \cite{FAS_channel 0.2}.}\label{sim:performance_versus_N_o}
\end{figure}
\subsubsection{Performance versus Active Ports Number}

In Fig.~\ref{sim:performance_versus_N_o}, the AD and CE performance are evaluated for varying numbers of active ports $ N_o $, with a fixed pilot length $ G=200 $, antenna length constant $ M=64$, $ K_a=50$ active users, and $ \mathrm{SNR}=-15 $ dB, under a Rician factor $ K_r=0 $ (NLOS).

All algorithms exhibit a sharp decline in ADE as the number of active ports increases. The greedy-based algorithms demonstrate faster decline in ADE with increasing $ N_o $, attributable to the additional observations. However, the proposed Algorithm II provides much enhanced CE NMSE through all varying $N_o$. The NMSE of the proposed Algorithm II  decreases rapidly, reaching approximately 0.085, which is much lower than that of all benchmarks reducing NMSE by 59.12\%. The NMSE floor of the proposed algorithm II derives from the fixed level of SNR and merely increasing the number of receiving ports does not reduce the interference level.

The performance improvement, driven by the exploitation of angular information, is advantageous and provides a strategy to address the potential performance floor issue discussed in Section~\ref{inherent_problem}. In contrast, the proposed Algorithm I and the conventional algorithms display different trends. For Algorithm I, the NMSE remains relatively stable at lower $ N_o $, such as $ N_o \le 12 $. However, as $ N_o $ increases to higher values, its performance slightly degrades and then remains steady after $N_o=52$. This intriguing behavior is likely due to the increased correlation among ports which can de understood as the issue of minimum port spacing. 

In \eqref{eq:21}, all ports are assumed to be independent, which does not hold with dense array within compact space. For FAS, optimizing port spacing is critical, though it is beyond the scope of this paper. When the gap between ports becomes too narrow, the independence among ports may be undermined which in turn perplexes the prior PDF modeling and thus impairs the performance of proposed algorithm I. Nevertheless, this does not significantly impair the performance of the proposed algorithms, provided the parameters are appropriately configured.

\begin{figure}[!t]
	\centering
	\includegraphics[width=\columnwidth]{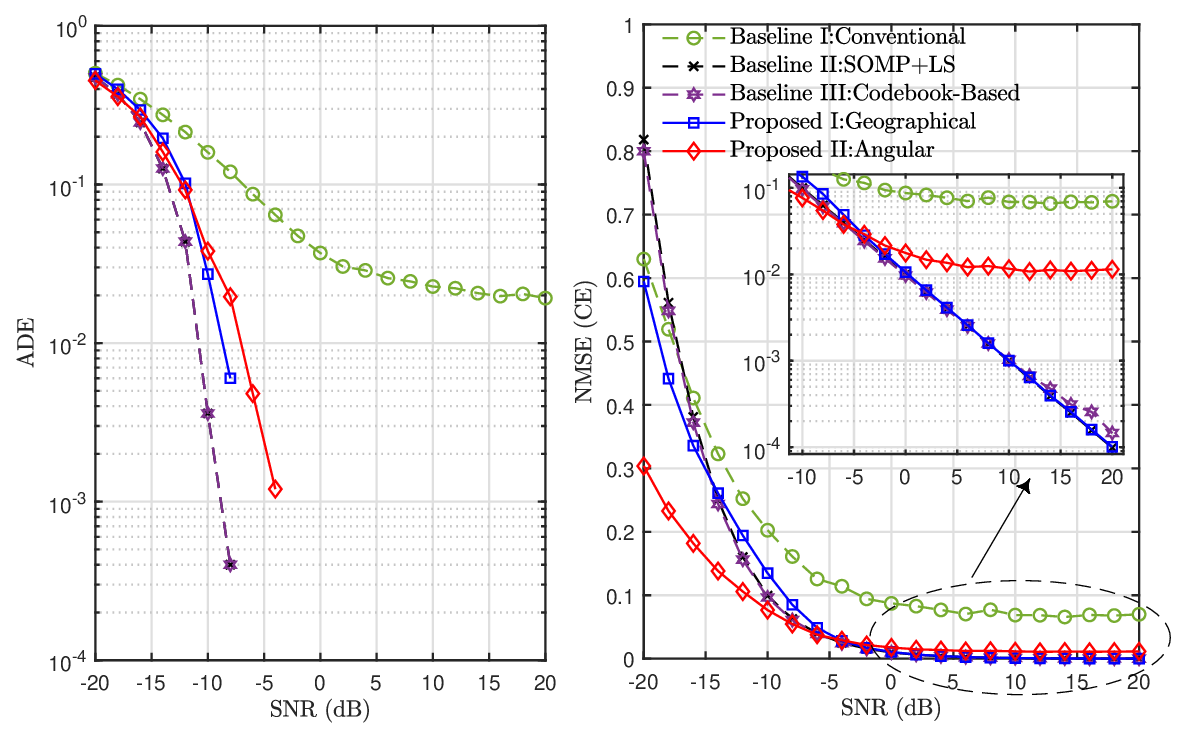}
	\caption{Illustration of ADE (left) and CE NMSE (right) of different algorithms versus SNR (dB) with antenna length constant $M=64$, pilot length $G=150$, $K_a=50$ active users and $N_o=16$ active ports. Rician factors are $K_r=0$ (NLOS). Performance baselines in comparison are conventional EM-AMP \cite{EM-AMP1}, SOMP+LS \cite{CE5} and AoA codebook-based \cite{FAS_channel 0.2}.}\label{sim:performance_nmse_ade}
\end{figure}
\subsubsection{Performance versus SNR (dB)}
Furthermore, it is demonstrated that the NMSE floor of the proposed algorithm II is influenced by the SNR level. Fig.~\ref{sim:performance_nmse_ade} depicts the ADE and NMSE (CE) performance across varying SNR (dB) and active port counts $ N_o =16$, with fixed parameters: antenna length $ M=64 $, pilot length $ G=150 $, $ K_a=50 $ active users, and Rician factor $ K_r=0 $ (NLOS).

In general, all algorithms exhibit improved ADE and CE performance as SNR increases, with the proposed Algorithm II achieving the highest estimation precision among them under low SNR region (before -10 dB) and with the proposed Algorithm I achieving similar performance to LS yet with reduced complexity. Though the conventional EM-AMP manifests a ADE performance floor, the proposed algorithms have a water-falling ADE performance akin to LS and greedy-based solutions.

\begin{figure}[!t]
	\centering
	\includegraphics[width=\columnwidth]{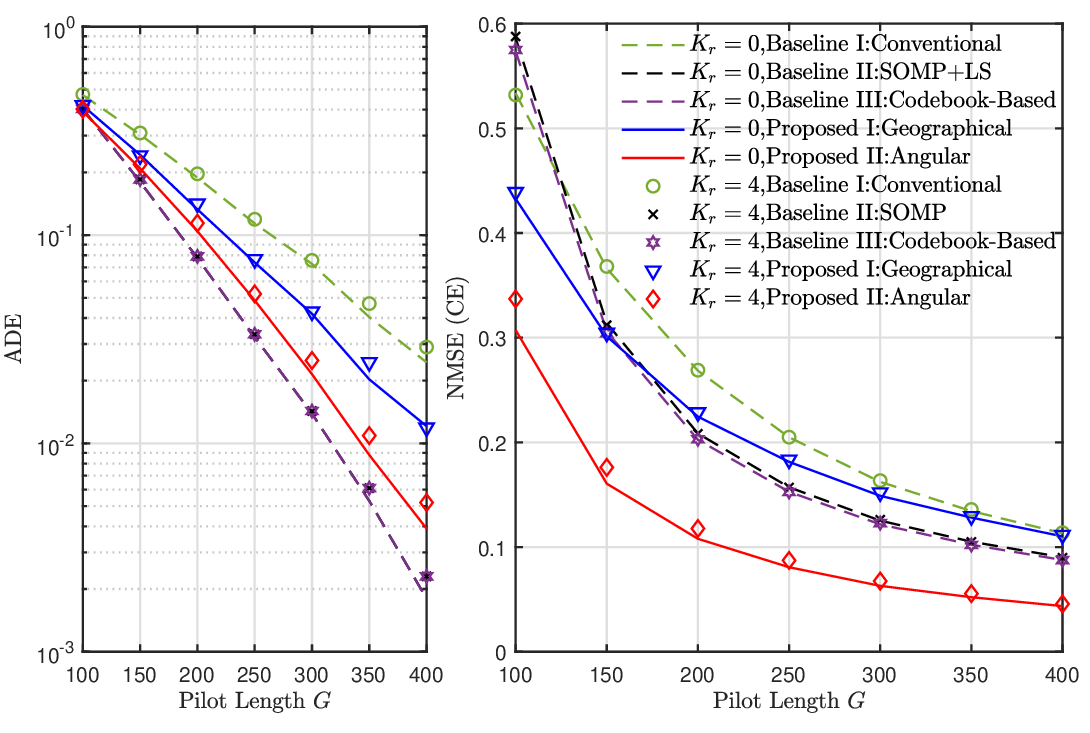}
	\caption{Illustration of ADE (left) and CE NMSE (right) of algorithms versus different pilot length $G$ with antenna length constant $M=64$, active port number $N_o=16$, $K_a=50$ active users and $\mathrm{SNR}=-15$ dB. Rician factors are $K_r=0$ (NLOS) and $K_r=4$ (LOS/NLOS). Performance baselines in comparison are conventional EM-AMP \cite{EM-AMP1}, SOMP+LS \cite{CE5} and AoA codebook-based \cite{FAS_channel 0.2}.}\label{sim:performance_versus_pilot_length}
\end{figure}
\subsubsection{Performance versus Pilot Length}

Pilot length $ G $ is a critical parameter for uplink training. Accordingly, Fig.~\ref{sim:performance_versus_pilot_length} illustrates the ADE and CE NMSE performance under varying $ G $, with fixed parameters: antenna length $ M=64 $, active port number $ N_o=16 $, $ K_a=50 $ active users, and $ \mathrm{SNR}=-15 $ dB, for Rician factors $ K_r=0 $ (NLOS) and $ K_r=4 $ (LOS/NLOS).

Consistent with prior findings, the Rician factor has minimal impact on the performance of all algorithms. Meanwhile, conventional EM-AMP exhibits the poorest AD performance, although its CE performance slightly surpasses that of SOMP before $G\le 150$. Furthermore, the proposed Algorithm II consistently delivers the best CE performance across all pilot length regions with 50\% NMSE reduction, highlighting its remarkable superiority in exploiting angular information.

\begin{figure}[!t]
	\centering
	\includegraphics[width=3.3 in]{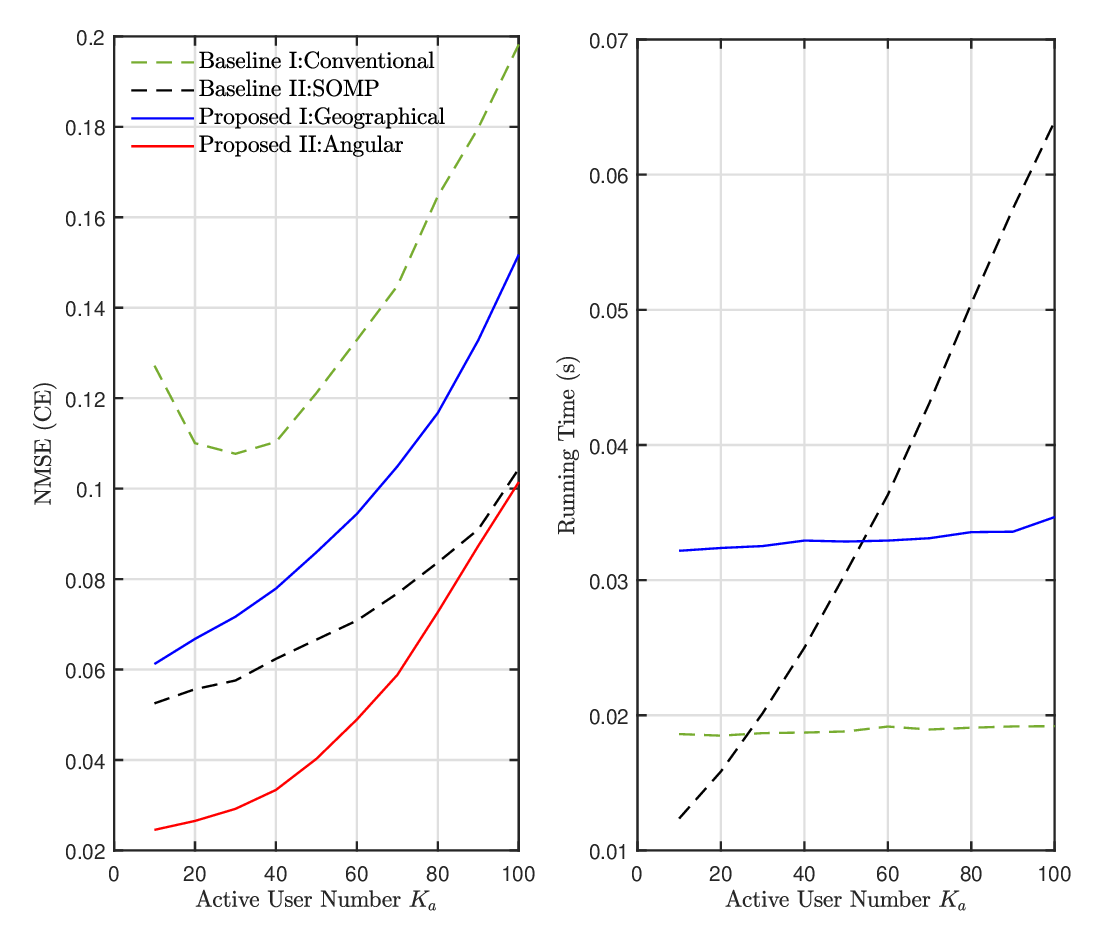}
	\caption{Illustration of running time (s) and the CE performance under different number of active users with antenna length constant $M=64$, $N_o=16$ active ports and $G=200$ pilot length with $\mathrm{SNR}=-10$ dB and AMP iteration upperbound $T_{\max}=5$.}\label{sim:running time}
\end{figure}
\subsubsection{Computational Complexity}
In this part, we illustrate CE performance under different number of active users with antenna length constant $M=64$, $N_o=16$ active ports and $G=200$ pilot length with $\mathrm{SNR}=-10$ dB and AMP iteration upperbound $T_{\max}=5$. The configurations of the simulator are 13th Gen Intel(R) Core(TM) i7-13700 (2.10 GHz), 32.0 GB RAM, Windows 11-24H2 with MATLAB R2024b.

For CE NMSE, the conventional EM-AMP exhibits a fluctuating tendency, likely due to a limited iteration upper bound, preventing performance convergence to a stable level. Regarding running time, with only a few users, the greedy-based SOMP demonstrates shorter running time, reflecting lower computational complexity. However, as the number of active users increases, the AMP-based algorithm outperforms SOMP+LS, as its complexity is determined by the codebook size, while SOMP+LS's complexity scales with the activity level. The simulation results align with the complexity analyses. Consequently, the AMP-based algorithm is better suited for massive connectivity in FAS scenarios.



\section{Conclusion}\label{conclusion}
In this work, we introduce the EM-AMP framework for CSI acquisition in FAS. We derive and propose update rules for two variants of the EM-AMP framework that leverage geographical and angular information within signals, achieving significant performance improvements in estimation precision and convergence speed. Notably, these variants exhibit favorable computational complexity in large activity regions compared to existing methods. We also provide an analytical explanation for the performance floor observed in existing methods. Furthermore, we analytically derive and empirically verify the estimation error deviations with and without angular information, emphasizing on the importance of angular information. 

Overall, this work demonstrates that AMP-based solutions for FAS CSI acquisition offer a well-balanced approach, characterized by low-complexity efficiency, feasible implementation, and signal model flexibility. In future, further exploration of AMP-based variants tailored to different channel or signal models will be considered.
\balance

\newpage

%
%
%
%

\vfill

\end{document}